\documentclass[useAMS,usenatbib]{mnras}
\pdfoutput=1

\usepackage{amsmath,amssymb}
\usepackage{caption}
\usepackage{subfig}
\usepackage{graphicx}
\usepackage{natbib}

\newcommand{\spose}[1]{\hbox to 0pt{#1\hss}}
\newcommand{\lta}{\mathrel{\spose{\lower 3pt\hbox{$\mathchar"218$}}
     \raise 2.0pt\hbox{$\mathchar"13C$}}}
\newcommand{\gta}{\mathrel{\spose{\lower 3pt\hbox{$\mathchar"218$}}
     \raise 2.0pt\hbox{$\mathchar"13E$}}}

\newcommand{\bI}{\mathbf{I}}

\newcommand{\bw}{\mathbf{w}}
\newcommand{\bl}{\mathbf{l}}

\newcommand{\bX}{\mathbf{X}}
\newcommand{\bF}{\mathbf{F}}

\newcommand{\msun}{M_\odot}
\newcommand{\kpc}{\,\mbox{kpc}}

\defcitealias{Weinberg:15a}{Paper 1}
\defcitealias{Weinberg:15b}{Paper 2}
\defcitealias{Weinberg:15c}{Paper 3}

\title[Computational KAM II]{Practical application of KAM theory to
  galactic dynamics: II. Application to weakly chaotic orbits in
  barred galaxies}

\author[M.~D.~Weinberg]{Martin D. Weinberg\thanks{E-mail:
    weinberg@astro.umass.edu} \\
  Department of Astronomy\\ University of Massachusetts, Amherst
  MA 01003-9305, USA }

\begin{document} 

\label{firstpage}

\pagerange{\pageref{firstpage}--\pageref{lastpage}} \pubyear{2015}

\maketitle

\begin{abstract}
  Owing to the pioneering work of Contopoulos, a strongly barred
  galaxy is known to have irregular orbits in the vicinity of the
  bar. By definition, irregular orbits can not be represented by
  action-angle tori everywhere in phase space.  This thwarts
  perturbation theory and complicates our understanding of their role
  in galaxy structure and evolution.  This paper provides a
  qualitative introduction to a new method based on KAM theory for
  investigating the morphology of regular and irregular orbits based
  on direct computation of tori described in \citet{Weinberg:15a} and
  applies it to a galaxy disc bar.  Using this method, we find that
  much of the phase space inside of the bar radius becomes chaotic for
  strong bars, excepting a small region in phase space between the ILR
  and corotation resonances for orbits of moderate ellipticity.  This
  helps explain the preponderance of moderately eccentric
  bar-supporting orbits as the bar strength increases.  This also
  suggests that bar strength may be limited by chaos!  The chaos
  results from stochastic layers that form around primary resonances
  owing to separatrix splitting. Most investigations of orbit
  regularity are performed using numerical computation of Lyapunov
  exponents or related indices. We show that Lyapunov exponents poorly
  diagnose the degree of stochasticity in this problem; the island
  structure in the stochastic sheaths allow orbit to change morphology
  while presenting anomalously small Lyapunov exponent values
  (i.e. weak chaos).  For example, a weakly chaotic orbit may appear
  to change its morphology spontaneously, while appearing regular
  except during the change itself.  The numerical KAM approach
  sensitively detects these dynamics and provides a model Hamiltonian
  for further investigation.  It may underpredict the number of broken
  tori for strong perturbations.
\end{abstract}

\begin{keywords}
  methods: numerical --- stars: kinematics and dynamics --- galaxies:
  kinematics and dynamics, formation, structure
\end{keywords}

\section{Introduction}
\label{sec:intro}

The dynamics of barred galaxies has influenced our understanding of
secular evolution and the importance of classical indeterminism or
\emph{chaos} over the last several decades
\citep[see][]{Contopoulos:02}.  In the standard non-linear dynamics
literature, chaos is often classified by the magnitude and character
of the exponential divergence of nearby trajectories in phase space.
If the chaos is strong, one may measure exponential divergence in at
least one dimension, and this may result in orbital diffusion in
otherwise conserved quantities of the motion.  Conversely, some orbits
may remain in a small region of phase confined by regions of
regularity, possibly with small or undetectably positive exponential
divergence; this chaos is often called `weak'.
\citet{Athanassoula2009,Athanassoula2009a,Manos2011,Contopoulos.Harsoula:2010,
  Kaufmann.Contopoulos:96, Patsis.etal:1997, Romero-Gomez.etal:2006,
  Romero-Gomez.etal:2007, Manos:2008, Tsoutsis.etal:2009,
  Brunetti.etal:2011, Bountis.etal:2012} have all discussed the role
of chaos in barred galaxies.  In addition, \citep{Cachucho.etal:2010,
  Giordano.Cincotta:2004, Manos.Machado:2014} have recently discussed
to the implications of irregularity in n-body simulations of barred
galaxies.  Clearly, it is well-established that chaos is pervasive in
the presence of strong patterns.  The work described here is motivated
by the desire to connect the onset of chaos to methods of classical
perturbation theory that underpins our theory of the strong patterns
themselves.  The term \emph{chaos} has broad context and appears to
have many definitions.  For the remainder of this paper, we will call
any orbit that can not be fully described by its conserved quantities
as \emph{irregular} or \emph{chaotic} synonymously, and we will refer
to periods of exponential sensitivity to initial conditions as
\emph{stochastic}.

The most commonly used technique to diagnose chaos in time-independent
potentials is Lyapunov exponent analysis.  These exponents, $\lambda$
measure the exponential divergence of two initially close initial
conditions $|\bX_1(0) - \bX_2(0)|={\cal O}(\epsilon)$ in phase space
in the form $|\bX_1(t) - \bX_2(t)| = \exp(\lambda t)|\bX_1(0) -
\bX_2(0)|$.  Lyapunov exponent analysis is robust: it can be applied
to any equations of motion regardless of complexity, although those
that can be linearised analytically have some computational advantage.
Let us consider the application of this method to a \emph{background}
model described by its full complement of actions, $\bI$, and a
perturbation which induces irregularity.  Chaotic orbits by nature
appear to diffuse through the previously invariant tori described by
$\bI$ and may be detected by positive Lyapunov exponents.  Other chaos
detection schemes, such as Fourier spectral analysis, short-term
Lyapunov analysis, and the recent the Generalised Alignment Index
(GALI) may be used in more general time-dependent systems, but we will
not consider these here.  Although many of these have been
productively and creatively used to investigate astronomical systems,
these exponents and indices can be difficult to link to the underlying
dynamical mechanisms relevant for astronomical problems.  For example,
Lyapunov exponents for weak chaos may be small or zero, but induced
change in the orbits may be qualitatively large and result in a
completely different orbit morphology.  Weak chaos appears to dominate
for barred systems and we will illustrate the effect on specific
orbits in section \ref{sec:results}.

In this paper, we will explore an alternative approach that
reconstructs the perturbed tori directly using a numerical technique
based on Kolmogorov-Arnold-Moser (hereafter, KAM) theory, described in
\citet[][Paper 1]{Weinberg:15a}.  The technique attempts to find a new
torus (regular orbit) using the constructive procedure outlined in
many proofs of the KAM theorem.  This method has several distinct
advantages: 1) it is computationally efficient relative to exhaustive
numerical integration of orbits; 2) it provides a statistical
assessment of the regular-orbit fraction in action space of the
unperturbed problem; this in turn, allows a direct connection with our
standard dynamical description of patterns and instabilities in
galactic dynamics; 3) it provides self-diagnostic information about
the broken torus, e.g. which resonances are responsible.  Our method
does not follow the KAM procedure precisely.  Since we are interested
in locating and diagnosing broken tori, we do not attempt to find
nearby action values that yield valid tori, but rather interpret the
failure in a statistical sense.  Although the proposed approach is
perturbative, it applies to both general time-dependent and
time-periodic perturbations.  \citetalias{Weinberg:15a} suggests that
the two tools together give a more complete picture than either one
alone.  We will see that this is true for the present application as
well.  Our main goal is to characterise the use of this new method on
a popular astrophysical dynamics problem and compare with the more
commonly used Lyapunov exponent method.

For completeness, we will begin with a short review of both Lyapunov
exponent analysis and the numerical KAM approach.  A more in-depth but
non-mathematical description of the KAM-motivated ideas can be found
in section \ref{sec:KAM}. Consider a general n-dimensional dynamical
system
\begin{equation}
\frac{d\bX}{dt} = \bF(\bX, t),
\label{eq:odeN}
\end{equation}
where $\bX$ is an $n$-dimensional vector (e.g. 6-dimensional phase
space) and $\bF$ is smooth continuous vector function (e.g. a
\emph{physical} force function).  Our initial condition is given by
$\bX(0)$ at $t=0$.  The $n$ Lyapunov exponents of the system are
defined by the rate of the logarithmic increase of the axes of an
infinitesimal sphere of states around $\bX(0)$.  To compute this rate,
one may use the tangent map given by the first-order expansion of
equation (\ref{eq:odeN}):
\begin{equation}
  \frac{\partial\delta\bX}{dt} = \left(\nabla\cdot\bF\right)\cdot\delta\bX
  \label{eq:DodeN}
\end{equation}
where $\nabla\cdot\bF$ is the $n\times n$ Jacobian matrix
$(\nabla\cdot\bF)_{ij}\equiv\partial F_i/\partial X_j$.  One of the
standard methods used to determine the full Lyapunov spectrum is due
to \citet{Shimada.Nagashima:1979} and \citet{Benettin.etal:1980} who
use a Gram-Schmidt reorthonormalisation procedure to identify the
principle axes of the exponentially diverging and converging
sphere. An explicit source code for computations based on this
procedure is given by \citet{Wolf.etal:1985}.  This method requires
solving the original system and its tangent space (as defined in
eq. \ref{eq:DodeN}) simultaneously with the equations of motion
(eq. \ref{eq:odeN}); that is, the augmented system has $n(n+1)$
coupled first-order equations. There are a number of more recent
developments that address some of the numerical difficulties inherent
in computing Lyapunov spectra.  The most serious of which is slow
convergence.  A strictly positive maximal Lyapunov exponent is often
considered as a definition of deterministic chaos, but it is difficult
to associate them with explicit dynamical quantities
\citep[e.g.][]{ChaosBook:2012}.  More narrowly, a positive maximal
Lyapunov exponent indicates exponential sensitivity to initial
conditions, a feature of an unstable \emph{and} an irregular system.

Despite its wide use, this approach has some intrinsic disadvantages
and interpretive limitations.  First, consider that the general
perturbed phase space is foliated by \emph{sheaths} of chaos around
the homoclinic trajectories primary resonances with the perturbation
(\citetalias{Weinberg:15a}, section 4.1 gives a simple example for a
toy model).  Direct numerical experiment has shown that trajectories
that are initially in the vicinity of a one of these chaotic sheaths
may later on appear regular owing to apparent capture by a regular
island; this stickiness may repeat throughout the computation
\citep{Harsoula.etal:2011}.  Secondly, owing to the diffusion of the
dynamical trajectory through the action space of the unperturbed phase
space, the action values of trajectory in the unperturbed phase space
may have drifted considerably before the time series is long enough to
compute the Lyapunov spectrum.  In other words, the long-term
converged Lyapunov exponent is telling us something about the evolving
dynamical system, but it is hard to relate this to the finite-time
situation typical of galactic dynamics.  Short-term Lyapunov analysis
and the various expansion and contraction indices may provide
additional useful information, but direct connection to dynamical
principle will remain challenging.  Finally, the stochastic layers
near primary resonances may still result in stochastic behaviour but
remain undetectable by Lyapunov exponents (i.e. weak chaos).  See
\citetalias{Weinberg:15a} for additional discussion.

These difficulties motivate the exploration of alternative algorithms
for the diagnosis of chaos dynamical systems.  In the approach to be
explored here, we take a constructive approach. Specifically, we begin
with the Hamilton-Jacobi (HJ) equation, which is the underlying
dynamical framework for action-angle variables.
\citetalias{Weinberg:15a} presents an algorithm for obtaining a
solution to the HJ equation that loosely follows the constructive
proof of the Kolmogorov-Arnold-Moser (KAM) theorem.  We interpret the
failure to find a solution of the perturbed HJ equation using the KAM
construction as indicative of a destroyed torus described by some
initial value $\bI$.  This is consistent with the results of explicit
integration presented in \citetalias{Weinberg:15a}.  In addition, the
numerical KAM (hereafter, nKAM) approach will give us expressions for
the perturbed by regular orbits that may then be used for studying
their implication for galactic structure.  For example, if we
determine the trajectories corresponding to the new albeit perturbed
tori, we can use perturbed tori as basis for new perturbation theory
(see section \ref{sec:summary} for additional discussion).  Moreover,
the nKAM procedure is generally must faster than Lyapunov analyses for
the same phase-space resolution.  Also, it is embarrassingly parallel
\citep{Foster:1995} for an external determined potential field.

In some ways, the method to be explored here suffers from the same
time-scale issue as Lyapunov-exponent computation; that is, tori
computation for a periodic system assumes an eternally-applied
perturbation, whereas most astronomical perturbations have finite
duration.  It is possible to use the nKAM method over a finite-time
disturbance using extended phase space or employing a Laplace
transform in the time domain.  This will be explored in a later
contribution.

In this paper, we will apply the nKAM approach to an idealised barred
galaxy where the bar is considered to be a perturbation to the
axisymmetric galactic disc.  We are aware that this is a tall order
for a perturbation theory, but it also provides a good test of its
domain of applicability.  Moreover, barred-galaxy dynamics is a mature
subject, largely explored with idealised analytic models, Poincar\'e
surface-of-section plots and Lyapunov spectra.  Our hope and goal,
here, is the gleaning of some alternative insight from the nKAM
approach.  Specifically, we will examine the tori broken by adding a
quadrupole perturbation to an initially axisymmetric exponential disc
(see \S\ref{sec:bar} for details).  The bar is modelled as quadrupole
perturbation, and for simplicity, we will consider two-dimensional
orbits at the midplane of the disc only.  More realistic bar models
and those with time varying parameters can also be studied using this
method.  As the bar amplitude increases, the number of broken tori
increases (\S\ref{sec:results}) as expected \citep{Manos:2008}.  In
the limit of a strong bar (i.e. a large fraction of the disc interior
to the bar is in the bar), many of the tori inside of the corotation
and in bands around ILR and corotation are destroyed, leaving a
smaller band of regular orbits of moderate eccentricity left to
support the bar.  To be sure, the trajectories for many of these
broken tori are not dramatically ``Brownian''; they appear to flip
between different approximate rosettes of varying eccentricity.
Presumably, some of these modes for these weakly chaotic orbits will
not be bar supporting all of the time even if they would be in their
unperturbed state.  Even though the underlying perturbation method may
operating beyond its domain of applicability, these results suggest
that chaos may play a role in limiting the growth and structure of
bars, and this is supported by the results of our Lyapunov exponent
analysis.

\section{Review of the theory}
\label{sec:KAM}

The technical details of the nKAM approach are described in
\citetalias{Weinberg:15a}. This section summarises the main ideas and
motivation without the technical and mathematical details.

We begin with the standard setting for the equations of motion in a
stellar system: classical Hamiltonian perturbation theory
\citep[e.g.][]{Binney.Tremaine:2008}.  The fundamental coordinate
system describing the regular orbits in a quasi-periodic system is the
action-angle system.  This system results from a canonical
transformation from the $2n$ coordinates and generalised momenta that
separate the gravitational potential to those that leave the
Hamiltonian a function of $n$ new generalised momenta only.  There are
$n=3$ dimensions for galaxies, although we will consider the
restriction to $n=2$ for our planar examples here.  The equation
defining implicit solution for the generator of the canonical
transformation that yields the action-angle system is the HJ equation
\citep[e.g.][]{Goldstein.etal:02}.  Each \emph{regular} orbit has a
fixed action vector, $\bI$, and a corresponding angle vector, $\bw$,
whose values linearly increase with time.  Geometrically, the $\bI$
vector defines a torus in phase space that becomes uniformly sampled
by the trajectory in time (except for a small set of degenerate
closed-orbit situations).

Let us assume that we have a solution of the HJ equation for some
\emph{background} Hamiltonian function that yields regular orbits
everywhere.  Then, let us add a perturbation with pattern speed
$\Omega_p$ and attempt to find a correction to the generating function
that solves the perturbed HJ equation.  If successful, we have a new
set of action-angle variables and new tori.  We get an algebraic
solution for the new canonical transformation by solving the HJ
equation after a Fourier transform in angle variables.  The solution
has the form of trigonometric series whose arguments have the form
$\bl\cdot\bw$ where $\bl$ is a vector of integers, the Fourier
indices.  Formally, the coefficients of this series solution has
\emph{vanishing denominators}, indicating the evolution of the
specific action values at the point of vanishing is unbounded to
linear order.  The vanishing denominators have the form
$\sum_jl_j\Omega_j - m\Omega_p$ where $l_j$ is the Fourier index of
the $j$ angle variable, $m$ is most often the azimuthal angular
harmonic, the $\Omega_j=\partial H/\partial I_j$ are the unperturbed
frequencies of $\bw$ which follows from Hamilton's equations.

If each term in the series solution is considered to be independent of
the others, the vanishing denominator problem is resolvable.  To see
this, let us restrict our attention to only term that generates a
particular vanishing denominator.  The associated non-linear problem
has a well-defined solution: it is that of a pendulum!  The
one-dimensional pendulum represents the resonant degree of freedom.
As individual orbits exchange angular momentum, the collective changes
in actions causes the perturbation to evolve by changing its pattern
speed or amplitude, and this drives the perturbation the past the
resonance for each susceptible trajectory.  If we apply the averaging
theorem to isolate the evolution to the specific degree of freedom
implied by the vanishing denominator, and self-consistently include
these changes in the overall solution, we get a prescription for
secular evolution.  Dynamical friction is the classic example of
secular evolution of this form \citep{Tremaine.Weinberg:84}.  This
theory is very well developed in celestial mechanics where the
perturbation strength tends to be very small.

Clearly, traditional secular evolution is only piece of the full
dynamical picture.  For example, the Kirkwood gaps and solar system
stability cannot be explained using the tools of secular evolution.
The nature and implications of the \emph{full} solution remained a
mystery until relatively recently in the history of astrophysical
dynamics.  Returning to the previous example, suppose that we refrain
from applying the phase averaging to isolate individual terms and
retain the whole series, vanishing denominators and all.  The simple
example of this is the two interacting resonances considered in
\citetalias{Weinberg:15a}, section 4.1.  A series of papers in the
early 1960s by Kolmogorov, Arnold and Moser
\citep[KAM][]{Kolmogorov:1954, Arnold:1963, Moser:1962, Moser:1966}
showed that solutions to the HJ equation nearly always exist as long
as we stay far enough away from the phase-space location of the
vanishing denominators.  For small perturbations, this is most of
phase space.  Note: this does not imply that the vanishing
denominators are of no consequence (see previous paragraph!) but
rather that the dynamics that destroy the tori do not destroy them
everywhere in the action-space neighbourhood of the initial orbit.
The general message of KAM theory is: a significant fraction of tori
remain after applying a perturbation as long as the amplitude is
sufficiently small.

Much of the elegant mathematical machinery for solar system dynamics
is difficult to apply to problems galactic dynamics for several
reasons.  First, the perturbations are not those of point masses but
structurally extended with one or more scale lengths.  In celestial
mechanics, expansion variables that induce rapidly
(i.e. exponentially) converging Fourier series are possible.  For
example, \citet{Laskar:1990} shows that series solutions are practical
for planetary perturbations where the controlling amplitude is the
solar mass ratio and multiplicative terms proportional to powers of
eccentricity and inclination. In galactic dynamics, the perturbations
are most often globally extended and the series do not converge as
quickly.  Secondly, the frequency space is simply related to
coordinate system in celestial dynamics, that is, the unperturbed
frequencies are equal and depend only on semi-major axis.  This not so
in galactic dynamics where the unperturbed forces are generally
non-central and the frequencies and actions must be computed
numerically.  Finally, in the decades that followed the publication of
the KAM ideas, researchers further elucidated the implication of the
KAM theory for practical dynamical systems, studying ever larger
perturbations and the implications for the structure of real-world
phase space.  A wide variety of numerical experiments suggests that
the converse of the KAM theorem qualitatively holds: that is, the KAM
tori can be destroyed by increasing the strength of the perturbation.
This point is especially relevant to galactic dynamics where
perturbations in the form of satellite encounters, non-axisymmetric
disturbances (bars and arms) and structural deformations (dark halos
and flat discs) are strong perturbations.

Our goal here is to use the nKAM approach to identify circumstances
where a large fraction of tori are destroyed and investigate the
implication of these broken tori for galaxy evolution.  This study
pushes the dynamics beyond that of secular perturbation theory by
examining the effect of many terms in the Fourier series solution
together. That is, we do not isolate and solve each term in the
Fourier series separately but include their mutual influence.  That
is, we allow may resonances to interact simultaneously.  However, to
make a connection with the familiar insights from classical
perturbation theory, we will focus on a particular term whose
magnitude is amplified by its proximity to the locus in phase space
where the denominator vanishes even though all resonant terms remain
in the solution.  Following the standard conventions, call this the
\emph{primary} resonance and the remaining multiple resonances
\emph{secondary} resonances.  Here, our target application is disc
bars considered as a perturbation to the axisymmetric equilibrium.
There are three strong primary resonances for a bisymmetric $m=2$
harmonic perturber: 1) the corotation resonance characterised by an
orbit moving at the same azimuthal frequency as the perturbation and
inner and outer Lindblad resonances, characterised by the two radial
oscillation for every azimuthal oscillation in the frame rotating with
the bar pattern.  The primary resonances in the artificial absence of
secondary resonances are characterised by closed orbits in this
rotating frame.  When isolated to the one-dimensional resonant degree
of freedom, this special trajectory that connects the unstable
equilibria and separates the rotation from the libration regions is
the \emph{homoclinic} orbit or \emph{separatrix}.

Owing to the significant strength of bar perturbations implied by
observations, there are often Fourier coefficients whose amplitudes
are not small relative to the primary resonance.  These secondary
resonances may have amplitudes close to that of the primary resonance,
making the distinction somewhat ambiguous.  Near the instability for
the primary resonance, the secondary terms, even when they are
relatively small, can induce significant changes in the trajectory. In
effect, the secondary terms noticeably change the trajectory that
would result in absence of these terms.  This results in a sheath of
stochastic behaviour around the original unstable homoclinic
trajectory.  For strong perturbations, the sheaths of the primary
resonance broadened by secondary resonances may result in large-scale
diffusion through phase space (i.e. strong chaos, see
\citealt{Zaslavsky:2007}).  Similar to the overall conclusions of
previous work, we do not find strong chaos to play a major role in
barred galaxy dynamics except, perhaps for nearly radial trajectories
and for near-circular orbits close to corotation.  Many authors have
studied this problem in some detail with a variety of novel numerical
techniques (see section \ref{sec:intro}).  For example,
\cite{Bountis.etal:2012} used a statistical approach to identify both
weakly and strongly chaotic orbits in a barred galaxy model.  Zotos,
Caranicolas and collaborators in a series have articles (see
e.g. \citealt{Caranicolas.Zotos:2013}) have used Lyapunov exponents
and related indicators to characterise chaos in a variety of
interesting circumstances.  It is not the goal of this paper to
explore the wide range of astronomical situations leading to chaos
but, rather, to attempt to connect the existence of chaos to standard
Hamiltonian perturbation theory and KAM theory.

In part, the proof of the KAM theorem rests on the rearrangement of
terms in the linearised HJ equation to yield a quadratically
converging recursive solution for the new Fourier coefficients.  The
details of the proof specify some specific conditions of smoothness
and distance in phase space from a resonance so that the iteration
converges.  The proof also exploits the exponential decrease of
coefficient amplitudes with increasing harmonic order, $\bl$.  The
insight from KAM is that one may truncate the expansion series at some
order maximal order $\bl_{max}$ and still recover the correct
dynamics.  This is borne out in convergence tests used to verify the
results described below in section \ref{sec:results}.  Satisfactory
convergence implies the existence of a torus\footnote{As mentioned
  previously, our goal is to assess the fraction of broken tori, not
  to show the existence of tori as in the KAM theorem.  To this end,
  we do not adjust the initial condition in the neighbourhood of the
  original actions to maintain the Diophantine condition that controls
  the small divisors.}.  In addition, the solution results from a
linearised equation, and it remains possible that this procedure
under- or over-predicts the existence of tori in practice.  Finally,
it is tempting to interpret the failure to find a torus as evidence
for a `broken' torus, and we will do this here.  We have applied the
nKAM procedure, the Poincar\'e surface-of-section (SOS) method, and
Lyapunov analysis to the same problem and compared the results
(e.g. section \ref{sec:results} and \citetalias{Weinberg:15a}, section
5).  Of course, these methods are sensitive to different aspects of
the dynamics, but taken together, the interpretation of `broken' tori
appears warranted.  The numerical details of applying this method are
described in \citetalias{Weinberg:15a}.

\begin{figure}
  \centering
  \includegraphics[width=0.48\textwidth]{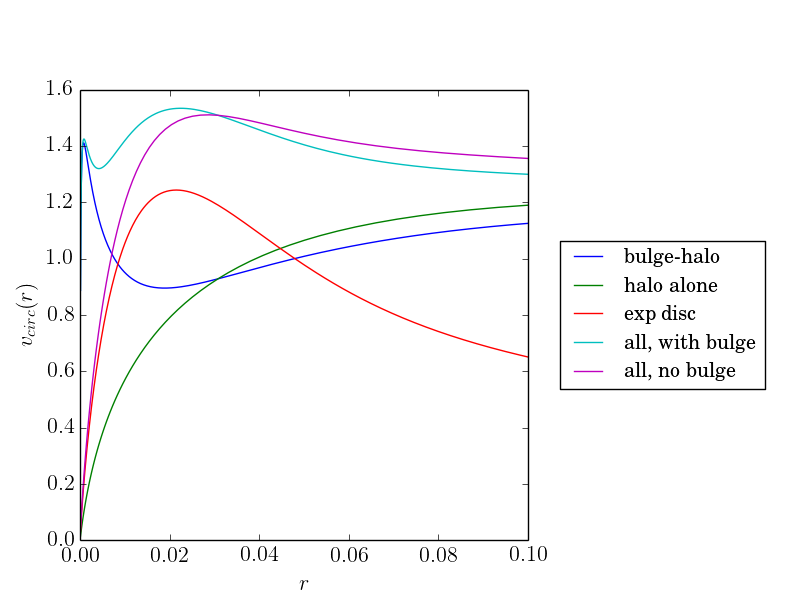}
  \caption{Circular velocity curve for halo and bulge (blue), the
    bulge-free halo model (green), the exponential disc (red), the
    combined model including the bulge (cyan), and the combined model
    without a bulge (magenta).  The bulge is constructed to provide an
    approximately flat rotation curve in the inner galaxy.  The
    bulge-free rotation curve rises inside of approximately two disc
    scale lengths.}
  \label{fig:rotcurve}
\end{figure}

\section{Galaxy model and bar perturbation}
\label{sec:bar}

We adopt the following set of $\Lambda$CDM units: the virial radius is
the unit length scale and the mass of the dark matter halo inside the
virial radius is unity\footnote{In $\Lambda$CDM, one uses the uniform
  spherical collapse model to define radius of collapse, $r_{\rm
    vir}$, for region of specified density enhancement.  Specifically,
  the virial mass, $M_{\rm vir}$, is the mass contained within a
  radius $r_{\rm vir}$ inside of which the mean interior density is
  $\Delta$ times the critical density $\rho_c$:
  \begin{equation*}
    \int_0^{r_{\rm vir}} r^2 {\rm d}r\ \rho(r) =
    {\Delta\over 3} \rho_c r_{\rm vir}^3
  \end{equation*}
  where $\rho(<r_{\rm vir}) = \Delta \rho_c$ is the halo's average
  density within that radius.  Cosmological n-body simulations are
  used to calibrate $\Delta$.  The virial radius may be related to
  point within which the material obeys the virial relation and
  external to which the mass is still collapsing onto the object.}.
Scaled to the Milky Way, the unit mass is approximately $10^{12}\msun$
and the unit radius is $300\kpc$.  We model the bulge and spheroid by
two NFW-like \citep[][NFW]{Navarro.Frenk.ea:97} profiles.  To combine
the dark-matter halo and bulge models, we ``leave room'' for the inner
cusp by adding a core to the dark matter halo by adopting a modified
NFW profile of the form
\begin{equation}
  \rho (r)=\frac{\rho_0 r_s^3}{\left(r + r_c\right)\left(r + r_s\right)^2}
\end{equation}
where $r_c$ is zero for the original NFW profile. We consider a model
with and without a bulge.  For the model with a bulge, the dark-matter
halo profile has a concentration of $c=15$ ($r_s=0.067$) and mass
$M_h=1$ inside of $r=1$, with $r_c=0.02$.  The bulge profile has a
concentration of $c=3000$ ($r_s=0.000{\bar 3}$) and a core radius
$r_c=1.{\bar 6}\times10^{-5}$.  This core radius limits the dynamic
range of the density inside of any radius of astronomical interest
here which simplifies some of the numerical computations.  For the
model without a bulge, the dark-matter halo profile has a
concentration of $c=15$ ($r_s=0.067$) and mass $M_h=1$ inside of
$r=1$, with $r_c=0.001$.

The two-dimensional disc is modelled by a thin exponential:
\begin{equation}
  \Sigma(R) = \frac{M_d}{2\pi a^2}e^{-R/a}
  \label{eq:discdens}
\end{equation}
The mass of the disc, $M_d$, was adjusted to produce an approximately
flat circular velocity curve from zero to ten disc scale lengths.
This mass is $M_d=0.04$ (approx. $4\times10^{10}\msun$ scaled to the
Milky Way).  The scale length is chosen to be $a=0.01$ units
(approx. $3\kpc$ scaled to the Milky Way).  The bulge mass is adjusted
to keep the inner profile flat rather than rising.  This gives a bulge
mass 0.02 units ($2\times10^{10}\msun$ scaled to the Milky Way) and a
B/D ratio of 0.5.  The bulge-free model has a rising rotation curve
out to several disc scale lengths. The circular velocity profile for
each component and the total for both models is shown in
Fig. \ref{fig:rotcurve}.  Motivated by both observations and the
results characterising many published n-body simulations, we chose the
bar length to be that of disk scale length, $a_1=a$.  Similarly, we
choose pattern speed $\Omega_p$ so that the bar rotates at the same
speed as a circular orbit at $R=a$.

Following previous contributions \citep[e.g.][]{Weinberg.Katz:02}, we
will model the bar as an ellipsoidal mass distribution.  The
gravitational potential for density stratified on ellipsoids is well
known.  Specifically, the gravitational potential internal to the
ellipsoid takes the form \citep[][pg. 52, eq. 93]{Chandrasekhar:69}
\begin{equation}
  V = \pi G a_1 a_2 a_3 \int^\infty_0\frac{du}{\Delta}
  \left[\Psi(1) - \Psi(m^2(u))\right]
  \label{eq:Vinner}
\end{equation}
where
\begin{equation}
  m^2(u) = \sum_{i=1}^3\frac{x_i^2}{a_i^2 + u}
\end{equation}
and
\begin{equation}
  \Psi(m^2) = \int_1^{m^2} dm^2\,\rho(m^2).
\end{equation}
The gravitational potential external to the ellipsoid has the form
\citep[][pg. 51, eq. 89]{Chandrasekhar:69}
\begin{equation}
  V = \pi G a_1 a_2 a_3 \int^\infty_\lambda\frac{du}{\Delta}
  \left[\Psi(1) - \Psi(m^2(u))\right].
  \label{eq:Vouter}
\end{equation}

We considered three different types of densities, power law,
\citet{Ferrers:1887}, and exponential:
\begin{eqnarray*}
  \rho(m^2) &=& \rho_0 
  \begin{cases} \left(1 - \log m^2\right) & \mbox{if }
    \gamma=-1 \\
    \left(1 - m^{2\gamma}\right) & \mbox{otherwise}
\end{cases} \\
  \rho(m^2) &=& \rho_0 \left(1 - m^2\right)^\gamma  \\
  \rho(m^2) &=& \rho_0 \left(e^{-1/\gamma} - e^{m^2/\gamma}\right)
\end{eqnarray*}
with $\gamma$ chosen to be physically sensible (i.e. $\gamma>=0$ for
the first two cases and $\gamma$ of order the disc scale length in the
final case).  The resulting quadrupole was similar with all three
models, and for ease, we choose to represent the bar with a
homogeneous ellipsoid with axes $a_1=a=0.01, a_2=0.005, a_3=0.001$.

\begin{figure}
\begin{center}
  \includegraphics[width=0.5\textwidth]{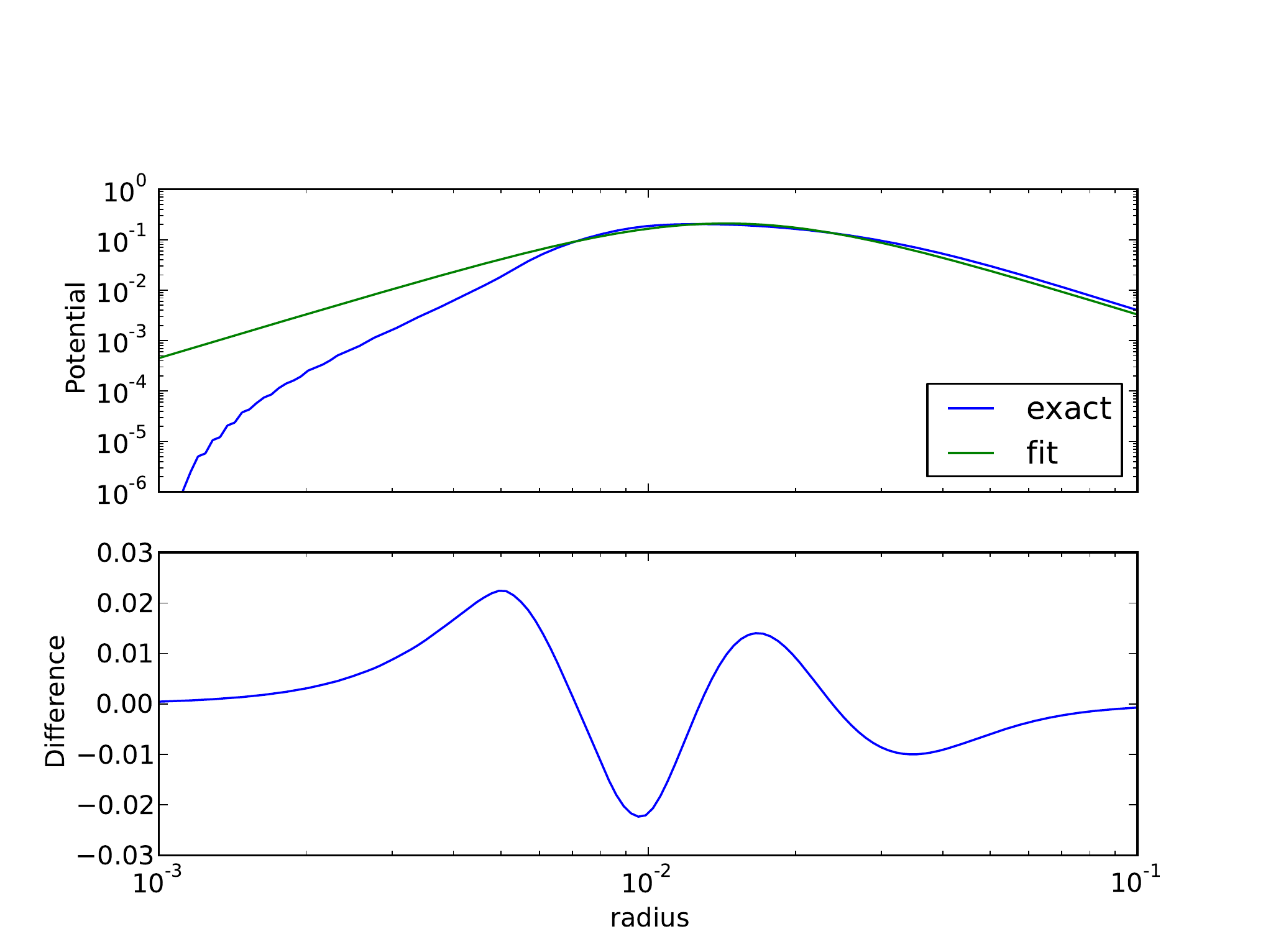}
\caption{Comparison of the exact quadrupole potential $U_{22}(R)$ from
  the homogeneous ellipsoid and the power-law-type fit
  (eq. \protect{\ref{eq:u22}}.  The upper panel shows compares the fit
in logarithmic units and lower panel shows the difference between the
fit and the exact quadrupole. \label{fig:u22fit}}
\end{center}
\end{figure}

To simplify the numerical computation, we fit the resulting quadrupole
to an analytic form which has the correct asymptotic power-law
behaviour based on the multiple expansion of Poisson's equation:
\begin{equation}
  U_{22}(r) = b_1 \frac{r^2}{1+\left(\frac{r}{b_5}\right)^5}
  \label{eq:u22}
\end{equation}
Fig. \ref{fig:u22fit} compares the fit from equation (\ref{eq:u22}) to
the quadrupole component of equations (\ref{eq:Vinner}) and
(\ref{eq:Vouter}).  The power-law fit systematically exceeds the true
value for small $r\ll0.01$ but otherwise captures the run of $U_{22}$
in the vicinity of the bar end at 0.01 quite well.

\begin{figure*}
  \centering
  \subfloat[$\epsilon=0.1$]{\includegraphics[width=0.49\textwidth]{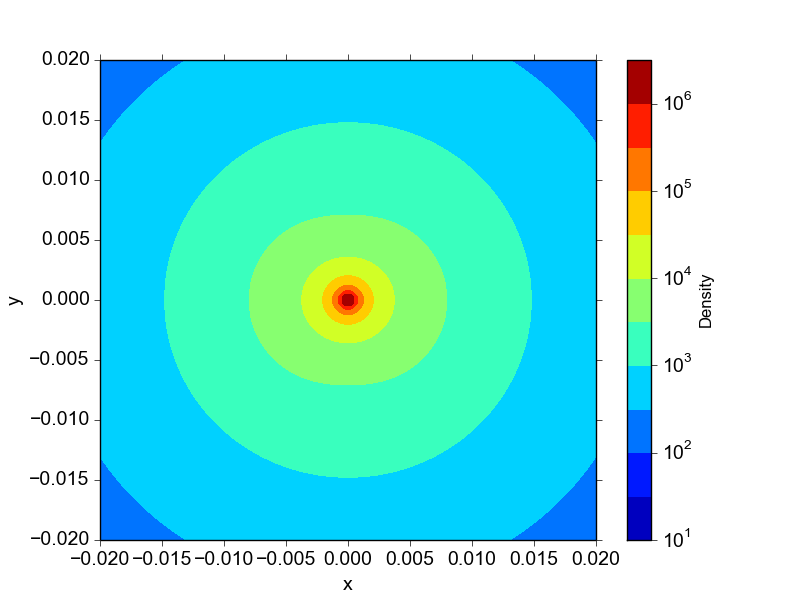}}
  \subfloat[$\epsilon=0.3$]{\includegraphics[width=0.49\textwidth]{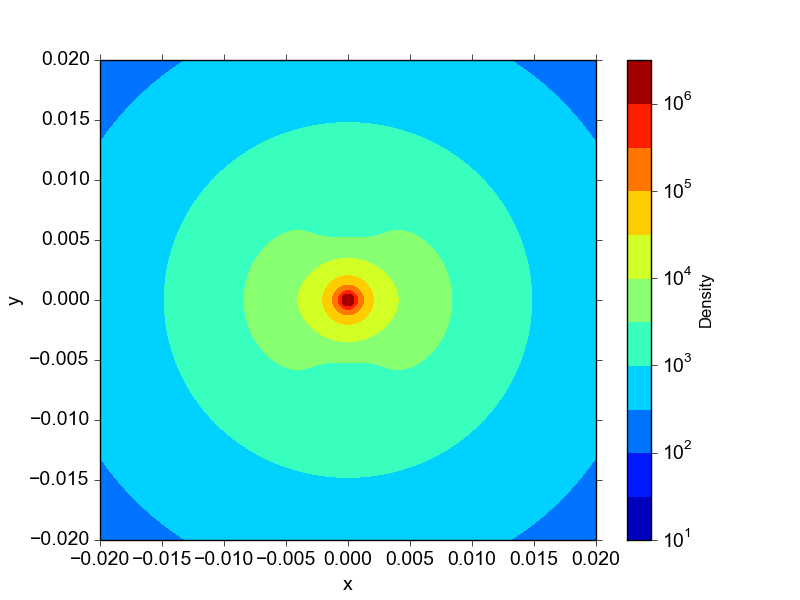}}

  \subfloat[$\epsilon=0.6$]{\includegraphics[width=0.49\textwidth]{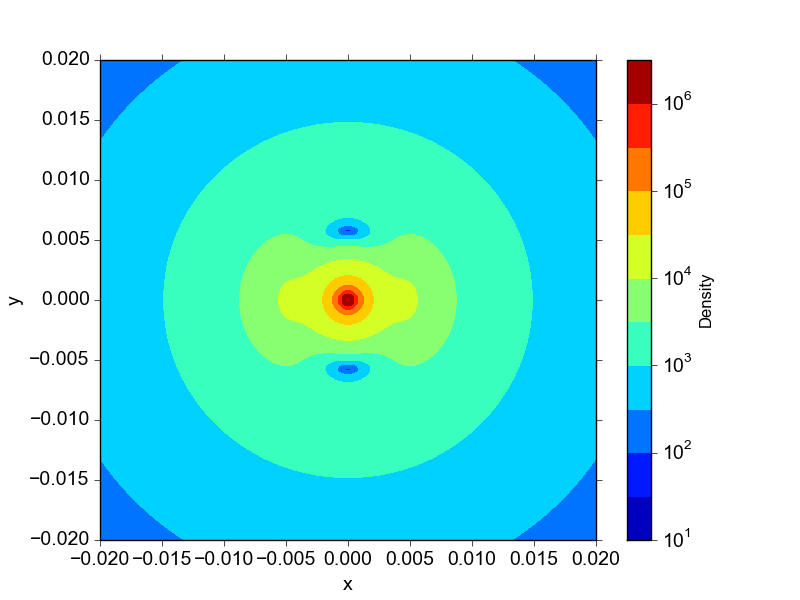}}
  \subfloat[$\epsilon=1.0$]{\includegraphics[width=0.49\textwidth]{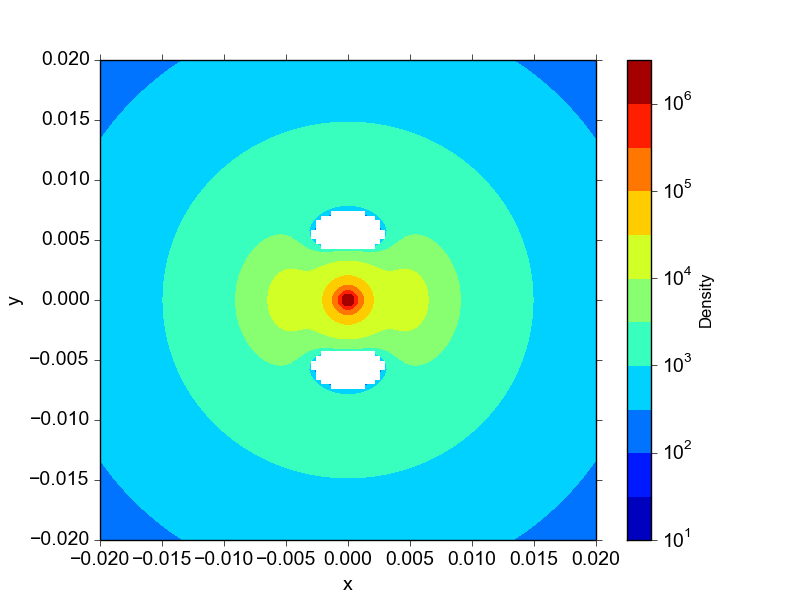}}
  \caption{Midplane density of the exponential disc with scale length
    $a=0.01$ (system units) including the bar perturbation for several
    amplitudes, $\epsilon$.  The bar scale length equals the disc
    scale length, $a=0.01$. \label{fig:bardens}}
\end{figure*}

\begin{figure*}
  \begin{center}
    \includegraphics[width=0.48\textwidth]{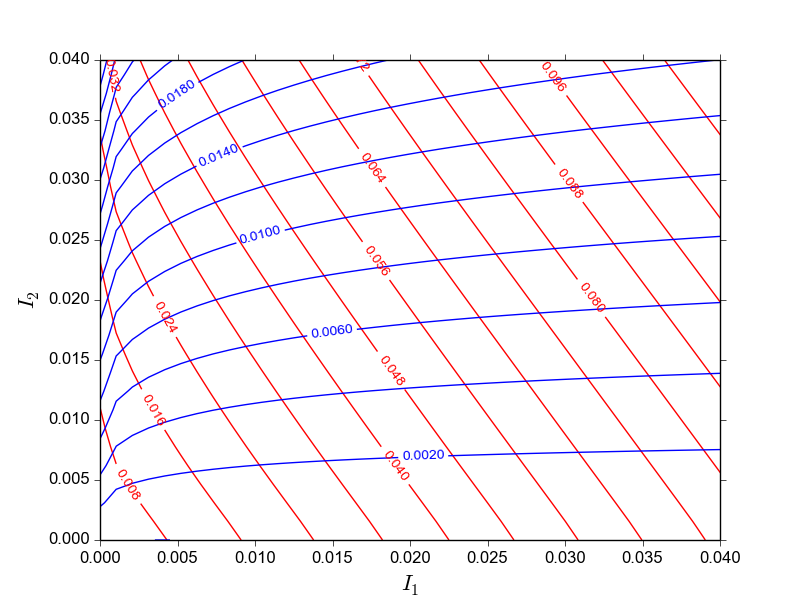}
    \includegraphics[width=0.48\textwidth]{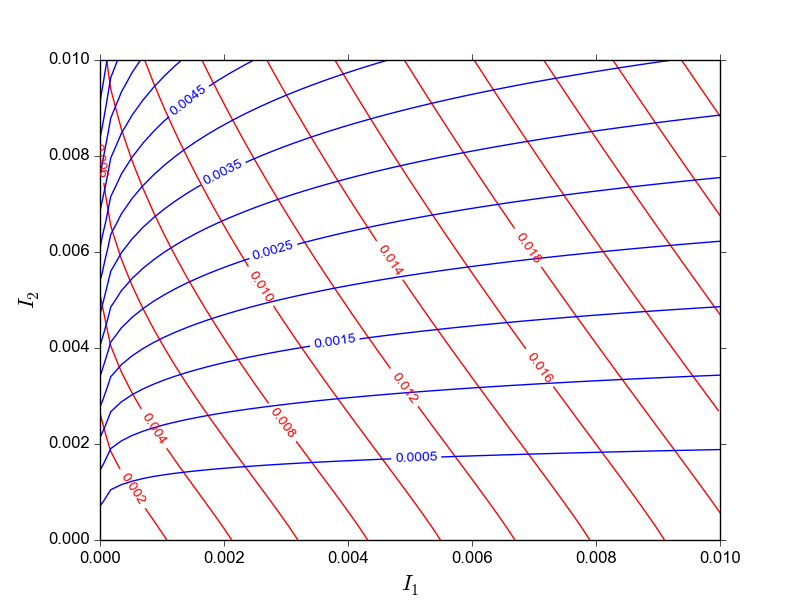}
    \caption{Mapping of action space to orbit turning points for the
      unperturbed model shown on large (left) and small (right) scale.
      Loci of inner turning points (blue) and outer turning points
      (red) are labelled in system units.  For reference, the bar
      length and disc scale length is 0.01 system units.
      \label{fig:apoperi}}
  \end{center}
\end{figure*}

\begin{figure}
  \includegraphics[width=0.5\textwidth]{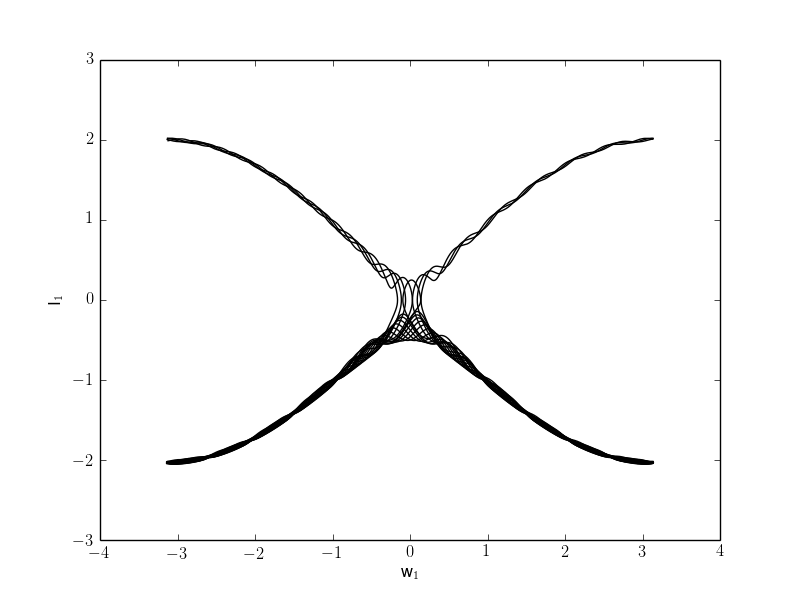}
  \caption{Twenty consecutive orbits from the two-resonance model from
    \protect{\citetalias{Weinberg:15a}} whose surface of section is
    illustrated section 4.1, Fig. 2 for an initial condition close to
    the homoclinic trajectory of the primary resonance.  As the
    trajectory approaches the unstable equilibrium of the primary
    resonance at $(I_1, w_1)=(0,0)$, the oscillations induced by the
    secondary resonance weave back and forth across the homoclinic
    trajectory, leading chaotic behaviour. \label{fig:turnstile} }
\end{figure}

\section{Broken tori and chaos}
\label{sec:results}

In this section, we use the nKAM method developed in
\citetalias{Weinberg:15a} and the bar perturbation described in
section \ref{sec:bar} to determine the location and examine the
morphology of broken tori.  The approach is that same as that
described for the Kepler example in \citetalias{Weinberg:15a}, section
5.  The unperturbed model is the entire axisymmetric component: bulge,
dark-matter halo and the exponential disc.  The quadrupole fit to the
triaxial ellipsoid representing the bar, described in the previous
section, is the perturbation. Thus, the perturbation amplitude only
affects the quadrupole strength.  We assume a constant pattern speed
$\Omega_p$ fixed to that of the circular orbit frequency at one disc
scale length.  As mentioned in section \ref{sec:KAM}, it is possible
to use the nKAM method to explore perturbations of finite duration;
these will be tackled in a later contribution.

Since the actions are the fundamental description of the unperturbed
phase space, we describe results of the nKAM procedure in the radial
action--azimuthal action plane (that is, the $I_1$--$I_2$ plane) for a
variety of bar quadrupole amplitudes.  The fiducial mass in the
triaxial ellipsoid that represents the bar is set equal disc mass
enclosed within the bar radius.  The resulting quadrupole amplitude is
scaled to the desired bar strength.  For visual comparison,
Fig. \ref{fig:bardens} shows the midplane density implied by the
quadrupole perturbation for four bar strengths spanning the range of
interest.  We use a logarithmic scaling for comparison with
astronomical images and to better illustrate the bar amplitude near
the ends of the bar.  The full-amplitude ($\epsilon=1$) profile bears
a close resemblance observed luminosity density for strong bars
\citep[e.g.][]{Gadotti.deSouza:2006}.  The negative-density
``dimples'' perpendicular to the bar that occurs for strongest
amplitude case is an artefacts of only including the quadrupole term;
higher-order multipole terms would fill in these regions perpendicular
to the bar's major axis.  The potential and force remain physical even
when dimple artefacts appear in the density profile.
Fig. \ref{fig:bardens} further suggests that perturbations with
$\epsilon\gtrsim 0.2$ are morphologically bar-like.

\begin{figure*}
  \centering
  \subfloat[$\epsilon=0.1$]{
    \includegraphics[width=0.48\textwidth]{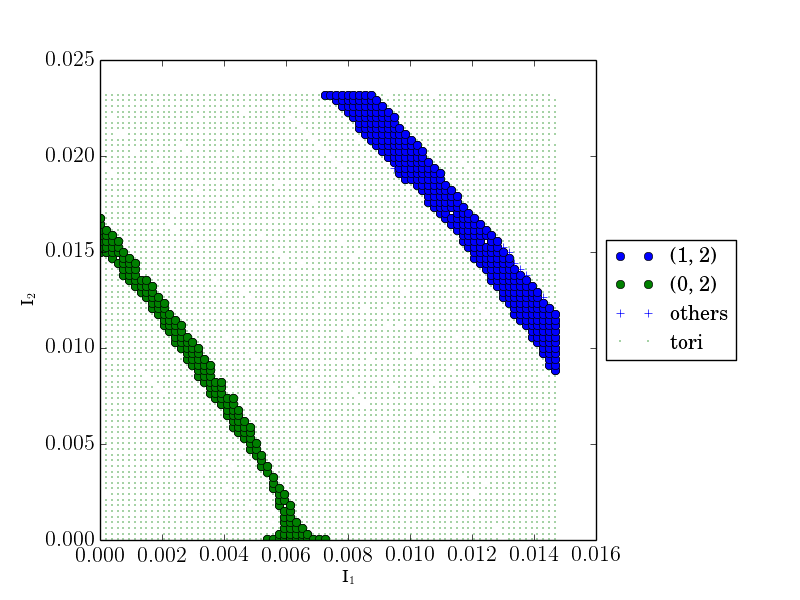}
    \label{fig:torie01k}
  }
  \subfloat[$\epsilon=0.3$]{
    \includegraphics[width=0.48\textwidth]{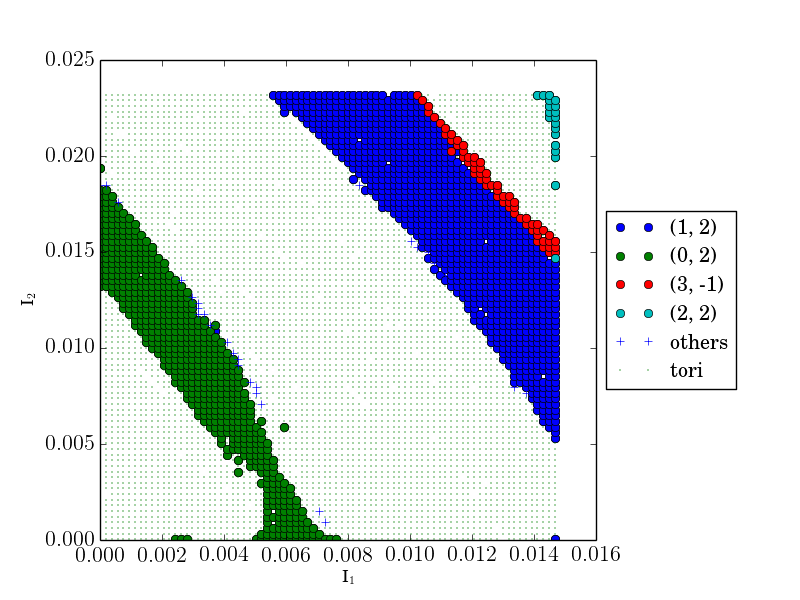}
    \label{fig:torie03k}
  }

  \subfloat[$\epsilon=0.6$]{
    \includegraphics[width=0.48\textwidth]{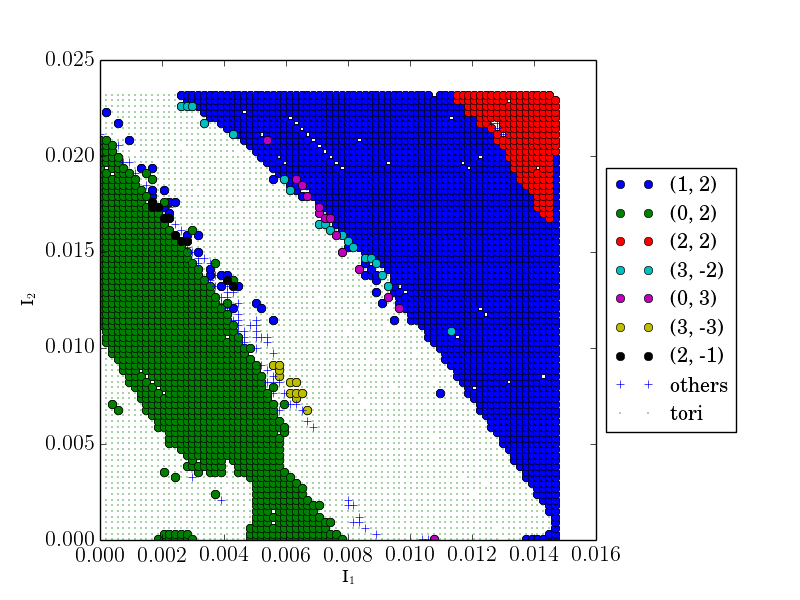}
    \label{fig:torie06k}
  }
  \subfloat[$\epsilon=1.0$]{
    \includegraphics[width=0.48\textwidth]{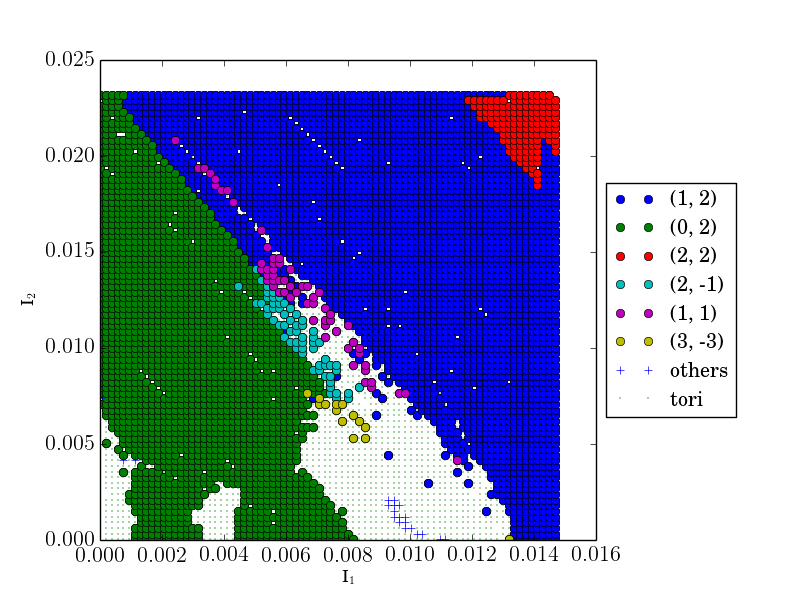}
    \label{fig:torie10k}
  }
  \caption{Broken (good) tori are indicated by filled circles (dots)
    as a function of the unperturbed values of radial action $I_1$ and
    azimuthal action $I_2$ for the bulge-free galaxy model. The broken
    tori are colour coded by the principle resonance in the form $(l_1,
    l_2)$ (legend).  For example $(-1,2)$ is the inner-Lindblad
    resonance and $(0, 2)$ is the corotation resonance. Initially
    circular orbits like along the $I_1=0$ vertical line and initially
    radial orbits like along the $I_2=0$ horizontal line.  The
    right-hand vertical axis describes the radius of the circular
    orbit ($I_1=0$) corresponding to each value of
    $I_2$.  \label{fig:torik} }
\end{figure*}

\begin{figure*}
  \centering
  \subfloat[$\epsilon=0.1$]{
    \includegraphics[width=0.48\textwidth]{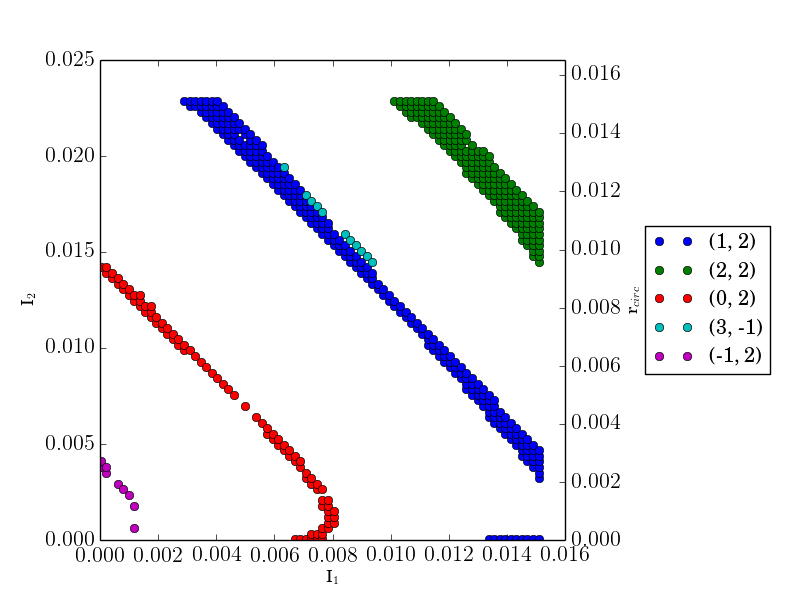}
    \label{fig:torie01}
  }
  \subfloat[$\epsilon=0.3$]{
    \includegraphics[width=0.48\textwidth]{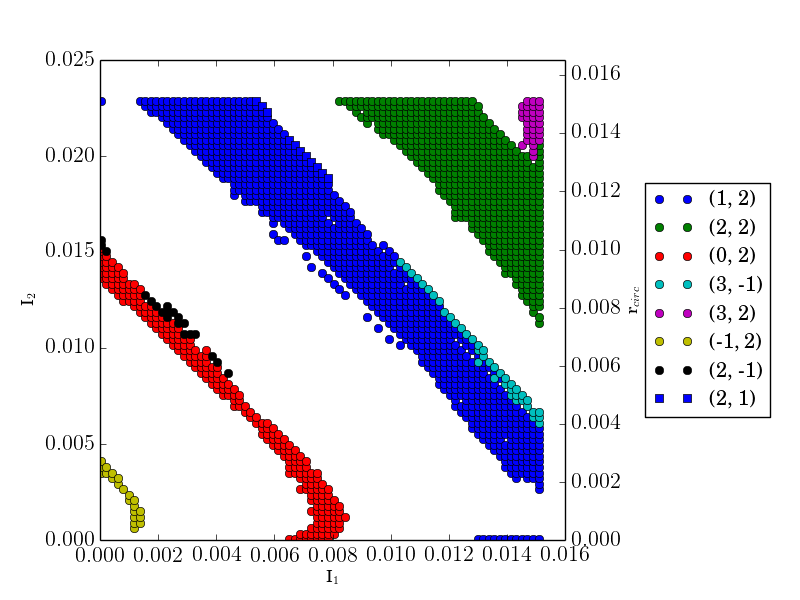}
    \label{fig:torie03}
  }

  \subfloat[$\epsilon=0.6$]{
    \includegraphics[width=0.48\textwidth]{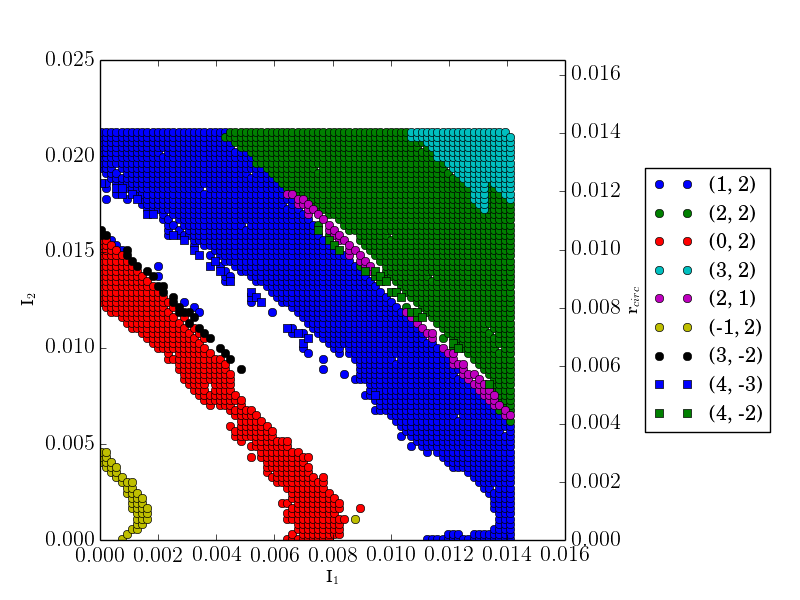}
    \label{fig:torie06}
  }
  \subfloat[$\epsilon=1.0$]{
    \includegraphics[width=0.48\textwidth]{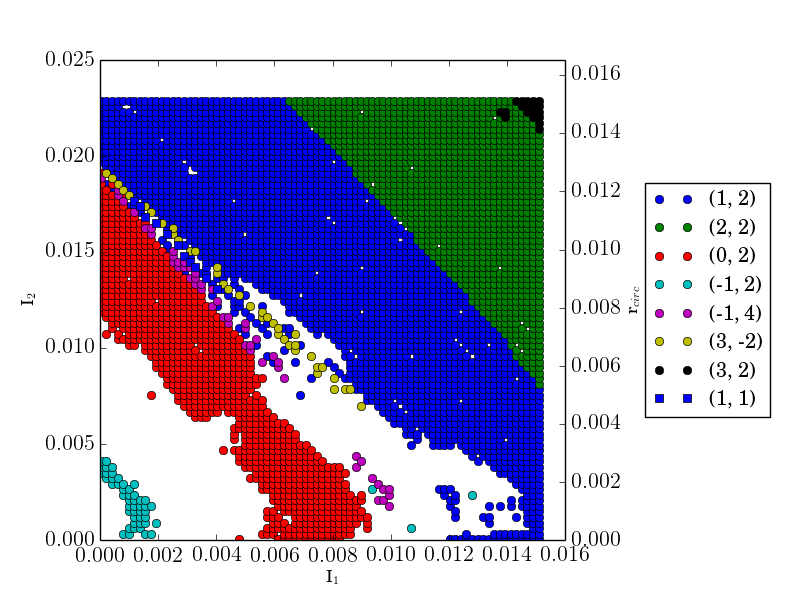}
    \label{fig:torie10}
  }
  \caption{As in Fig. \protect{\ref{fig:torik}} but for the galaxy model
      with a bulge and flat rotation curve.
      \label{fig:torie} }
\end{figure*}

The broken-tori action plots are shown in
Figs. \ref{fig:torie01k}--\ref{fig:torie10k} for the bulge-free,
rising rotation curve model and in
Figs. \ref{fig:torie01}--\ref{fig:torie10} for model including a bulge
with a relatively flat rotation curve.  The sequence of panels in each
figure corresponds to perturbation strength $\epsilon=0.1, 0.3, 0.6,
1.0$.  Each \emph{filled} point denotes a broken torus; the colour key
labels each by the term in the perturbation series (see
\citetalias{Weinberg:15a}, eq. 16) with maximum amplitude. The loci of
broken tori follow the low-order resonances, as expected.  For
example, the locus of broken tori at the lower-left in
Fig. \ref{fig:torie03k} follows the corotation resonance $(0, 2)$,
followed by the outer Lindblad resonance (OLR), $(1, 2)$, moving to
the upper right.  By examining the coefficients in the
generating-function series from \citetalias{Weinberg:15a}, eq. 16, we
can identify the secondary resonance or combination of multiple
secondary resonances implicated in splitting the separatrix of the
primary resonance.  In the case of the corotation resonance, the
secondary resonances are most often the inner Lindblad resonance (ILR,
$l_1=-1, l_2=2$), and vice versa for the primary ILR.  For
$\epsilon=0.1$ the relative strengths of the secondary-resonance
coefficients $\varpi_{\bl}$ of the $W_1$ series fall between 0.3\% and
3\%; these amplitudes seem to increase roughly linearly with
increasing $\epsilon$.  As the amplitude continues to increase, the
width of the region near the locus of corotation, OLR, and ILR
increase, starting to close the gap of unbroken tori between ILR and
corotation available to support the bar figure.  The main differences
between the two series, Figs. \ref{fig:torie01k}--\ref{fig:torie10k}
and Figs. \ref{fig:torie01}--\ref{fig:torie10}, are the locations and
existence of various resonances.  A flat rotation curve results in
more resonances near and around vicinity of the bar, including the ILR
at small radii.

We will see in section \ref{sec:results} below that the nKAM
computation appears to underestimate the measure of stochastic orbits
for mild eccentricity, typical of a disc, suggesting that a large
fraction of orbits in the bar region are stochastic at some level.
Indeed, computation of the SOS plots for in this region suggests most
of the orbits in this region with near circular orbits, $I_1\ll I_2$,
show some stochastic spreading.  In many but not all cases, these
irregular orbits appear to have inner and outer turning points
although the envelope of the trajectory vary with time. 
This calculation is not self consistent so the we have no way of
assessing whether the remaining unbroken tori are capable of
reproducing the imposed gravitational perturbation.  Nonetheless, the
nKAM results suggest that a large fraction of the phase space in the
bar vicinity is chaotic owing to bar itself.  This raises the
possibility that the bar amplitude and structure may be self-limited
by stochasticity.

\begin{figure*}
  \subfloat[$\epsilon=0.1$]{
    \includegraphics[width=0.48\textwidth]{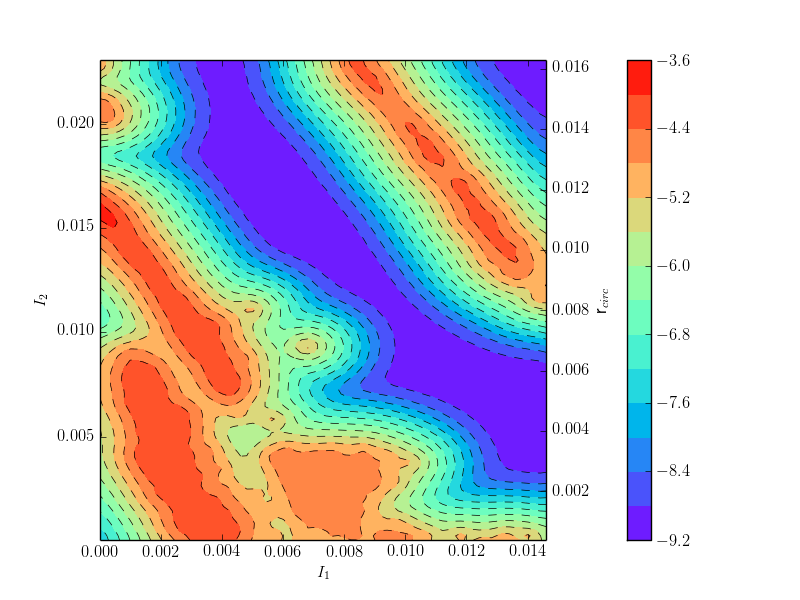}
    \label{fig:run42k2} 
  }
  \subfloat[$\epsilon=0.3$]{
    \includegraphics[width=0.48\textwidth]{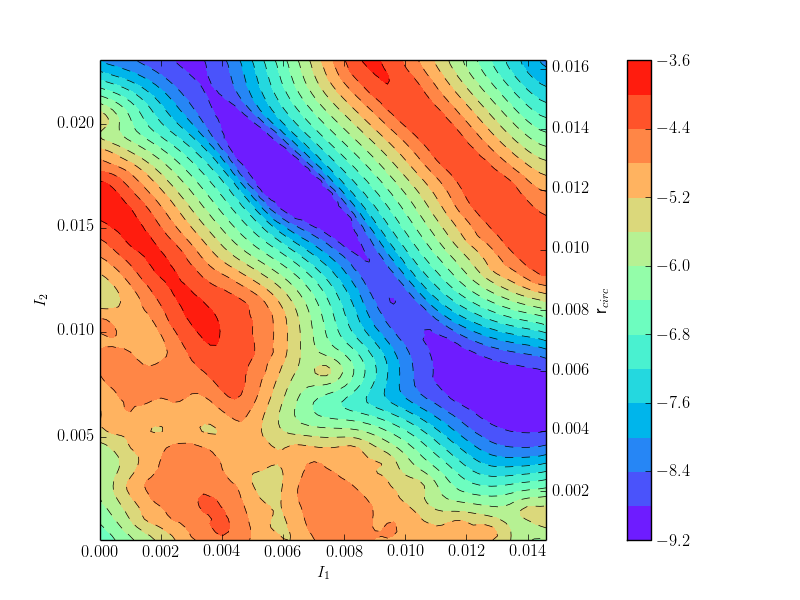}
    \label{fig:run42k3} 
  }

  \subfloat[$\epsilon=0.6$]{
    \includegraphics[width=0.48\textwidth]{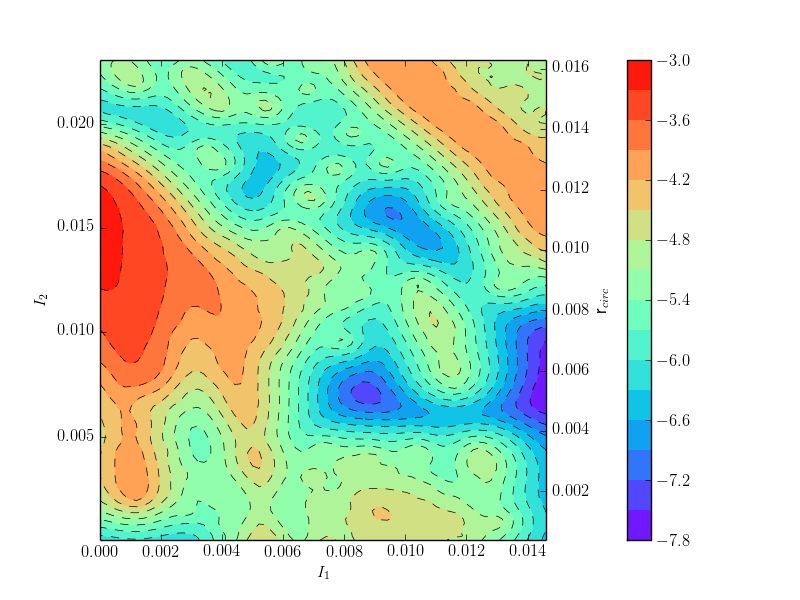}
    \label{fig:run42k4} 
  }
  \subfloat[$\epsilon=1.0$]{
    \includegraphics[width=0.48\textwidth]{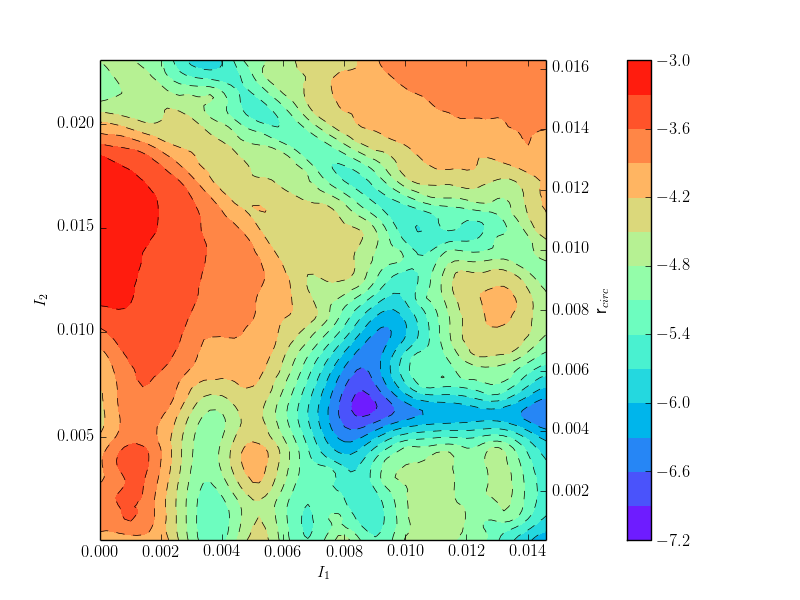}
    \label{fig:run42k5} 
  }
  \caption{Values of the  largest (positive) Lyapunov exponents for models
    with various perturbation amplitudes for the bulge-free galaxy
    model.  Positive Lyapunov exponents are determined by regression
    analysis to determine convergence with increasing time as described in
    \protect{\citetalias{Weinberg:15a}}, section 3.2.  The local estimate is
    computed by smoothing the ensemble onto a regular 
    grid. Contour values are logarithmically spaced; values of $(I_1,
    I_2)$ that are converging to zero are shown in white.
    \label{fig:lyapdensk} }
\end{figure*}

\begin{figure*}
  \subfloat[$\epsilon=0.1$]{
    \includegraphics[width=0.48\textwidth]{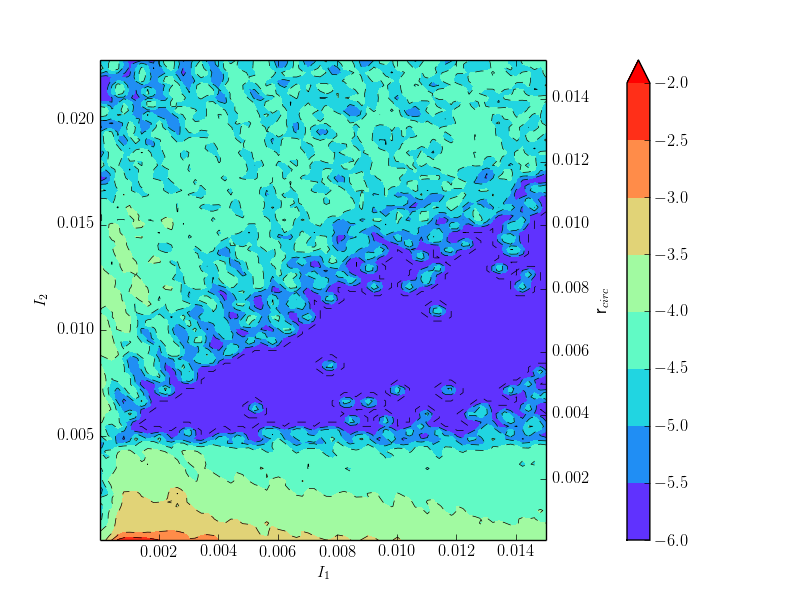}
    \label{fig:run42i2} 
  }
  \subfloat[$\epsilon=0.3$]{
    \includegraphics[width=0.48\textwidth]{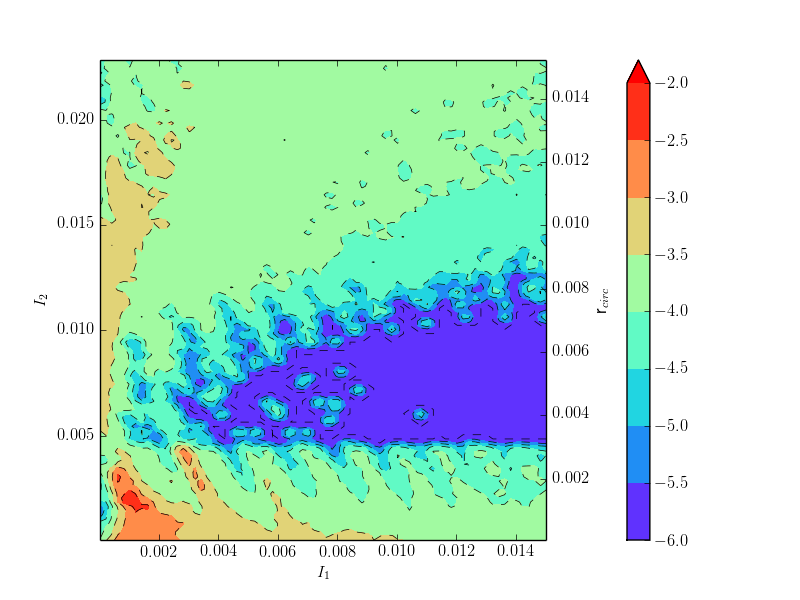}
    \label{fig:run42i3} 
  }

  \subfloat[$\epsilon=0.6$]{
    \includegraphics[width=0.48\textwidth]{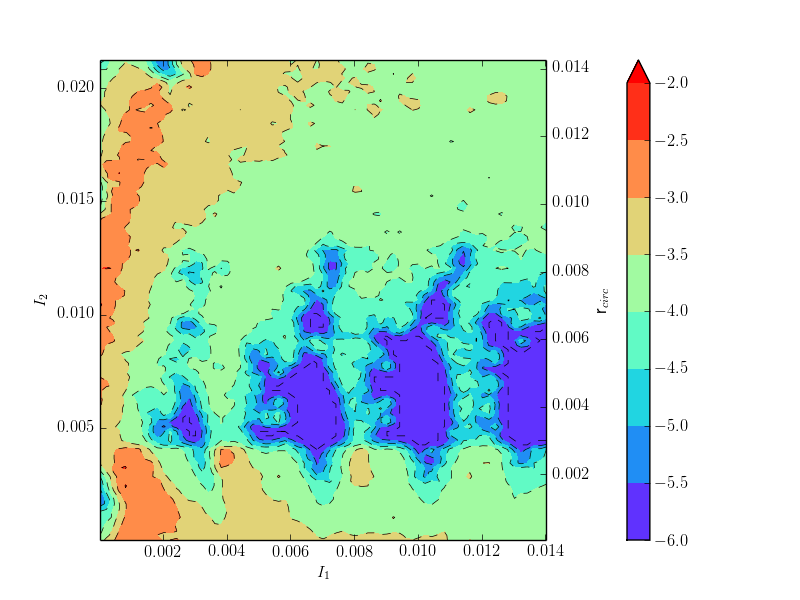}
    \label{fig:run42i4} 
  }
  \subfloat[$\epsilon=1.0$]{
    \includegraphics[width=0.48\textwidth]{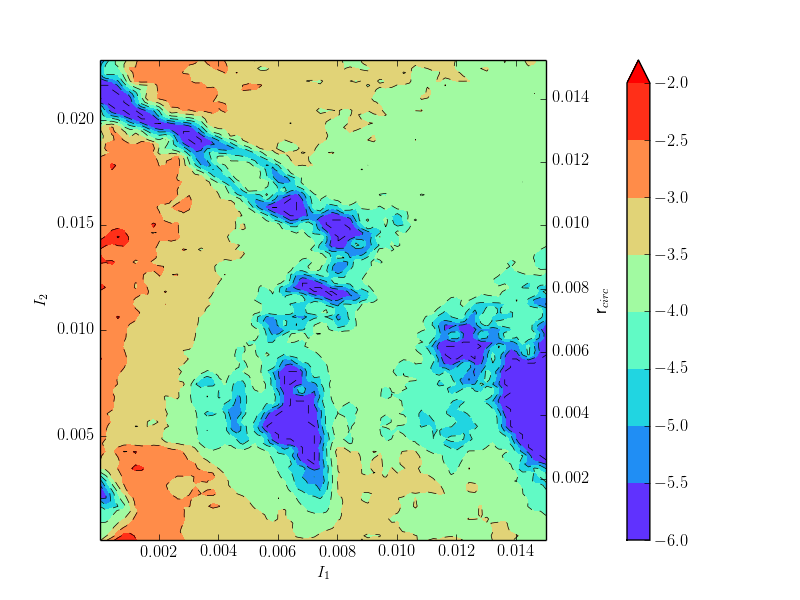}
    \label{fig:run42i5} 
  }
  \caption{As in Fig. \protect{\ref{fig:lyapdensk}} but for the galaxy
    model with a bulge and flat rotation curve.
    \label{fig:lyapdens} }
\end{figure*}

\begin{figure*}
  \centering
  \subfloat[regular zone, trajectory]{
    \includegraphics[width=0.48\textwidth]{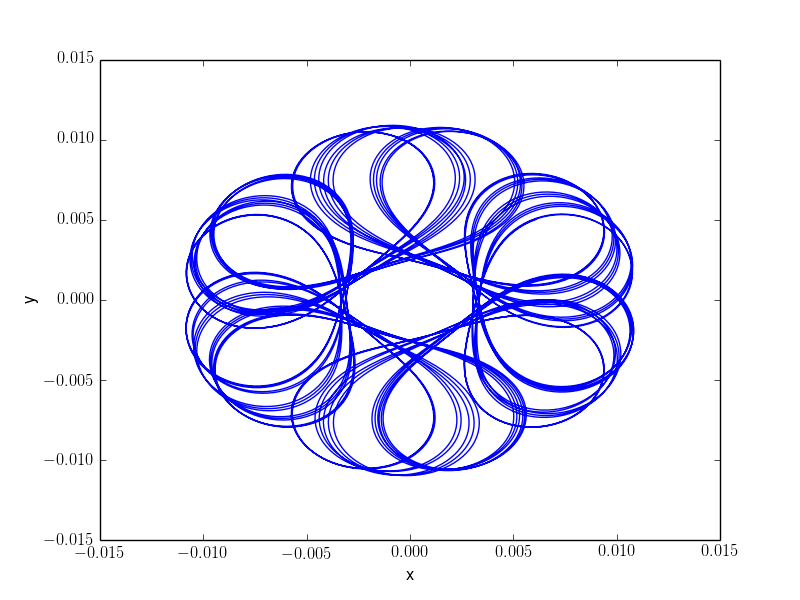}
    \label{fig:regular_orb}
  }
  \subfloat[regular zone, sos]{
    \includegraphics[width=0.48\textwidth]{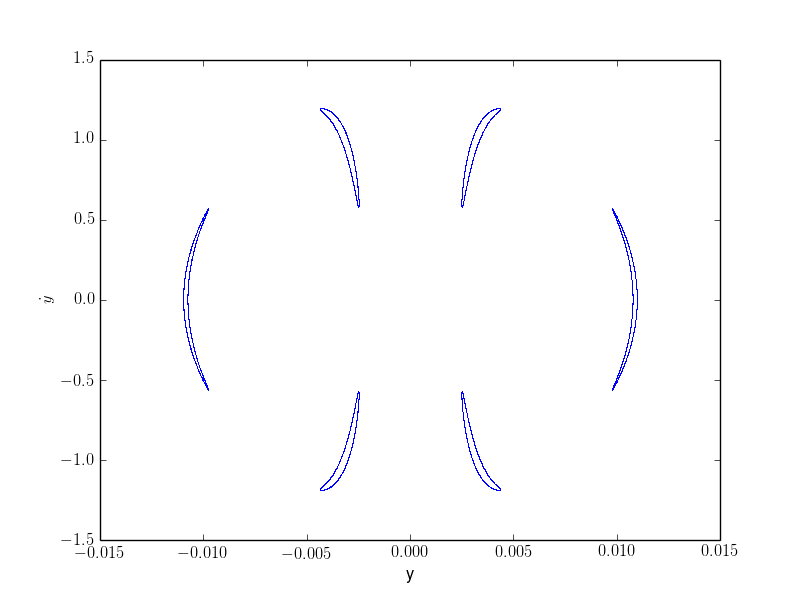}
    \label{fig:regular_sos}
  }

  \subfloat[chaotic zone, trajectory]{
    \includegraphics[width=0.48\textwidth]{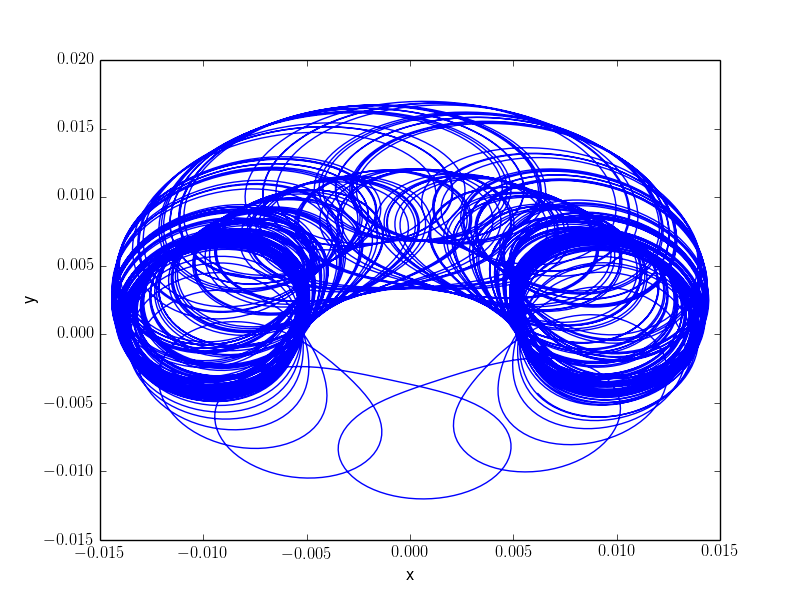}
    \label{fig:chaotic_orb}
  }
  \subfloat[chaotic zone, sos]{
    \includegraphics[width=0.48\textwidth]{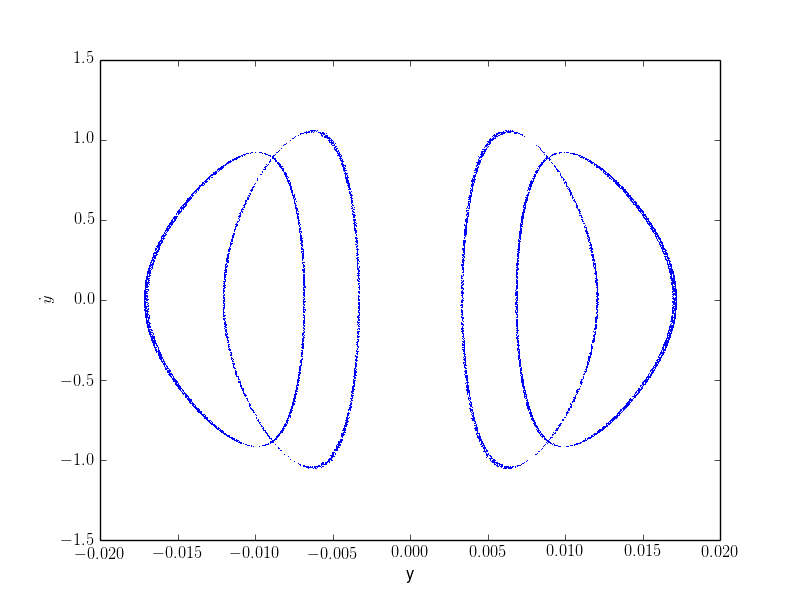}
    \label{fig:chaotic_sos}
  }
  \caption{Orbits shown in the frame rotating with the bar for
    $\epsilon=0.3$.  The regular orbit constructed by the nKAM method
    has $(I_1, I_2) = (0.0025, 0.007)$ and the chaotic orbit has
    $(I_1, I_2) = (0.0017, 0.011)$.  Both orbits have nearly the same
    Lyapunov exponent value: $\lambda\approx2\times10^{-5}$. The
    left-hand panels plot a short segment of the trajectory used to
    compute the two-sided SOS plots in the right-hand
    panels. \label{fig:test_orbits} }
\end{figure*}

We now compare the results of a Lyapunov exponent analysis with nKAM
predictions.  The presence of a bulge adds an additional interpretive
complication so we begin by describing the results for the bulge-free
galaxy model with a rising rotation curve.  The regions of positive
Lyapunov exponent for the amplitudes in
Figs. \ref{fig:torie01k}--\ref{fig:torie10k} are depicted in
Figs. \ref{fig:run42k2}--\ref{fig:run42k5}.  The locus of positive
exponents and broken tori coincide approximately for the corotation
and OLR resonances at all amplitudes.  As the bar amplitude increases,
corotation appears in positive Lyapunov values at high ($I_1\gtrsim
I_2$) and low ($I_1\ll I_2$) eccentricity.  Although the ILR does not
appear as a primary resonance, it \emph{is} a secondary resonance near
the inner edge of the region of broken tori corresponding to the
corotation resonance in all of
Figs. \ref{fig:torie01k}--\ref{fig:torie10k}.  At lower amplitude, the
ILR shows up as a distinct locus (Figs. \ref{fig:run42k2} and
\ref{fig:run42k3}).  At higher amplitude, regions of positive Lyapunov
exponent owing to corotation and ILR blend together in a single locus
(Figs. \ref{fig:run42k4} and \ref{fig:run42k5}).

For the flat rotation curve model, the regions of positive Lyapunov
exponent for the amplitudes in
Figs. \ref{fig:torie01}--\ref{fig:torie10} are depicted in
Figs. \ref{fig:run42i2}--\ref{fig:run42i5}. For all amplitudes, we
find a qualitative demarcation at $I_2\approx0.004$, corresponding to
orbits that move between the bulge and the disc (compare the rotation
curve contributions in Fig. \ref{fig:rotcurve} with the turning point
locations in Fig. \ref{fig:apoperi}).  Resonances between the driving
frequency of the bar and periods of control by the higher orbital
frequencies of the bulge lead to the striations in Lyapunov exponent
seen in these figures.  The nKAM convergence has been checked by
doubling and quadrupling the number of terms in the series, implying
that high-order resonances are responsible for the striation. As the
bar strength increases, the inner quadrupole, with a gravitational
potential $\propto r^2$, strongly affects the bulge dynamics.  The
perturbation causes the location of these high-frequency striations to
change location in action space with bar strength.  In essence, the
striations manifest in Figs. \ref{fig:run42i2}--\ref{fig:run42i5} are
a form of orbit ``shocking.''

The locus of positive exponents and broken tori coincide approximately
for the ILR at all amplitudes.  SOS plots confirm that these orbits
exhibit significant irregularity.  As the bar amplitude increases,
corotation appears in positive Lyapunov values at high ($I_1\gtrsim
I_2$) and low ($I_1\ll I_2$) eccentricity.  All in all, the overall
conclusions for both galaxy models are qualitatively the same:
increasing amplitude gives rise to an increasing fraction of
stochastic orbits in the bar region.  The galactic model with a flat
rotation curve has a larger zone of unbroken tori that are bar
supporting than the model with a rising rotation curve. For both
models, the nKAM analyses predict a region at moderate eccentricity
($I_1\lesssim I_2$) just below the chaotic corotation regions where
the orbits remain approximately regular in morphology and may support
the bar.  This zone of regularity decreases with increasing bar
amplitude.

A major advantage of the nKAM approach is that it naturally provides a
dynamical characterisation for the broken torus, i.e. hints to the
primary and secondary resonances involved.  On the other hand, a
broken torus predicted by the KAM method does not predict the degree
of exponential sensitivity.  However, empirically, a ratio of
secondary to primary amplitude of the generating function coefficients
$\varpi_{\bl}$ \citepalias[][eq. 16]{Weinberg:15a} of order 0.1 or
larger nearly always has a numerically measurable positive Lyapunov
exponent.  In essence, the secondary resonances break the forward and
backward time symmetry of the stable and unstable branches of the
resonant orbit.  Chaos is induced by rapidly oscillating intersections
of the perturbed stable and unstable manifolds in the sense of the
Birkhoff-Smale theorem \citep{Smale:65} that gives rise to exponential
sensitivity.  The resulting chaos is confined to a sheath surrounding
the original unperturbed homoclinic trajectory \citep[as
in][]{Holmes.Marsden:82}.  The dynamics of this behaviour was
considered in detail by \citet[][Chapter 4]{Wiggins:1990}.  It can be
easily seen numerically by plotting the solution of a direct numerical
integration for the many resonance system in the action-angle space of
the one-dimensional primary resonance near the homoclinic trajectory.
For example, Fig. \ref{fig:turnstile} shows a

The resulting motion for
broken tori of this type remains generally confined in phase space.
However, as the perturbation grows, the chaotic sheath can fill a
large fraction of the original libration zone, which may be large.  A
typical orbit in this sheath appears to quickly vary its eccentricity;
visually it presents as a rosette with turning points varying in
radius with time.

This insight helps us interpret the differences between the
approaches.  Recall that the Lyapunov value indicates the exponential
rate of divergence of two initial conditions.  However, this does not
describe the behaviour of the trajectory.  Qualitative classification
of orbital regularity by eye combined with the computation of SOS
plots suggests significant irregularity for exponents with $\lambda =
{\cal O}(10^{-3})$.  In many cases, the nKAM locus provides a more
sensitive indicator of separatrix disruption than the Lyapunov
exponent.  For example, Fig. \ref{fig:test_orbits} depicts the
trajectories and SOS plots for two orbits with the same value of $I_1$
and nearly the same value of Lyapunov exponent.  One orbit is in the
chaotic zone parented by corotation as predicted by the nKAM method,
and one orbit is between the ILR and corotation chaotic zones
[cf. Fig. \ref{fig:torie03}].  The orbit that is predicted to be
regular by nKAM shows qualitatively regular features, and this is
corroborated by the SOS plot.  The orbit that is predicted to be
chaotic by nKAM has two modes; the perturbation has broken the
original torus resulting into two loop-like trajectories that joined
at an unstable point.  This behaviour is typical of broken tori in the
corotation loci of Fig.  \ref{fig:torie} for $I_1\gta I_2$.  Broken
tori with $I_1\ll I_2$ exhibit more complex behaviour, presumably from
the superposition of effects from multiple secondary terms.  This
``mode switching'' explains the qualitatively `chaotic' appearance of
trajectory but the small measured Lyapunov exponent values.  Only near
the hyperbolic point of the islands in the stochastic sheath does the
exponential sensitivity to phase space play a strong role, and
therefore, the time-averaged Lyapunov exponent can remain anomalously
small in light of the qualitative stochastic behaviour of the
trajectory over many periods.

In summary, each of these two methods, nKAM and Lyapunov exponent
computation, reveal different aspects of irregular systems.  The nKAM
procedure identifies locations in phase space where influence of other
nearby or strong resonances are capable of inducing morphological
changes in near resonant orbits.  Lyapunov-exponent analysis attempts
to measure the rate of exponential sensitivity directly.  This
technique is good at discovering strong chaos but may miss subtle
cases where the mild break up of a resonance into several islands that
permits an orbit to spontaneously change morphologies after many
dynamical times.  In this later case, the trajectory is only
exponentially sensitive to its initial conditions near an unstable
point, and the time to transverse this critical region is a small
fraction of its orbital time.  While in this critical region, the
orbit may change from one nearly regular morphology to another.  In
other words, the characteristic time scale for exponentiation,
measured using the standard Lyapunov exponent algorithm, may be much
longer than an a Hubble time but the broken torus can have a
qualitatively different character owing to its tendency to switch
between morphologies.

\begin{figure*}
  \includegraphics[width=0.48\textwidth]{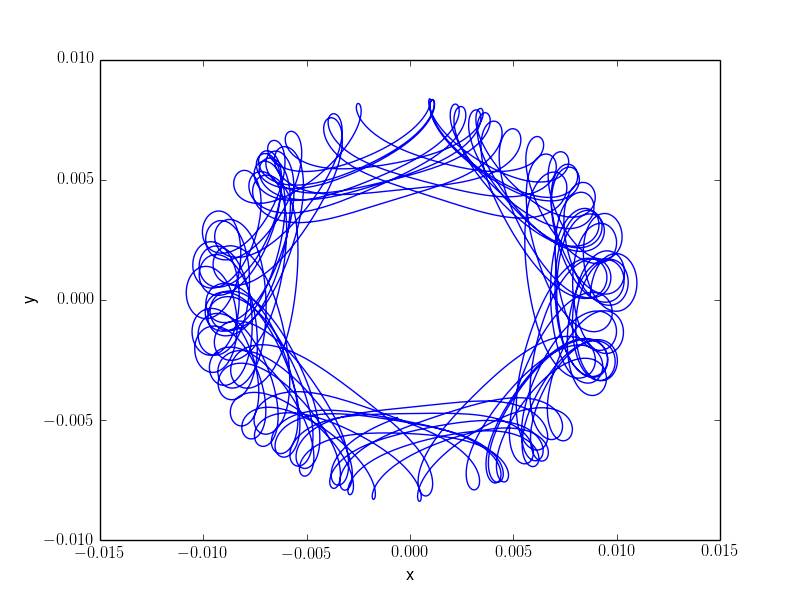}
  \includegraphics[width=0.48\textwidth]{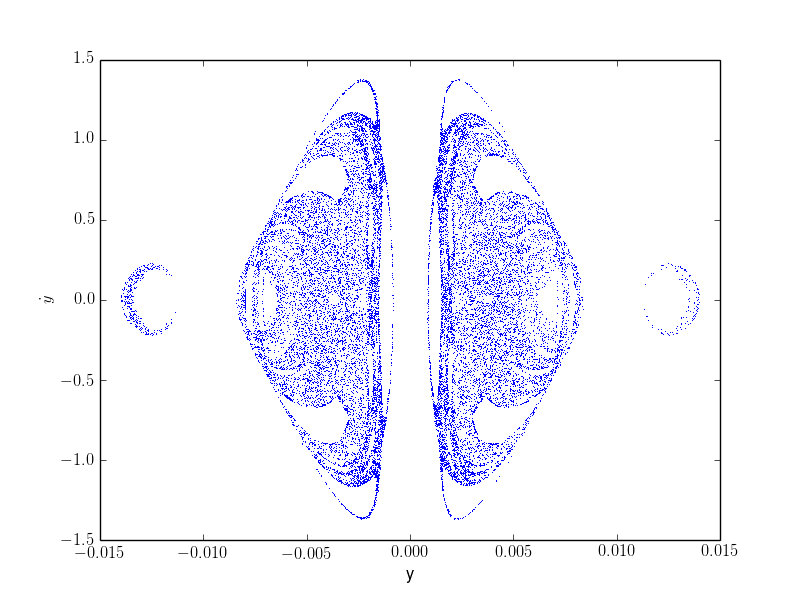}
  \caption{Orbit (left) and two-sided Poincar\'e SOS plot (right) for
    an initially nearly circular orbit with $(I_1, I_2)=(0.0004,
    0.011)$ at the middle of Lyapunov value plot for bar strength
    $\epsilon=0.3$ in Fig. \ref{fig:run42i3} but outside the chaotic
    zone in Fig. \ref{fig:torie03}.
    \label{fig:orb130} }
\end{figure*}

\begin{figure*}
  \includegraphics[width=0.48\textwidth]{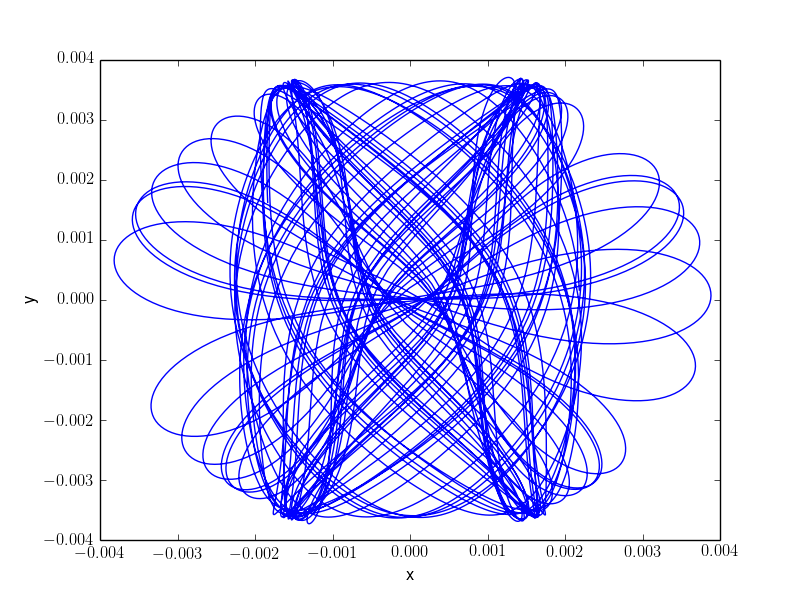}
  \includegraphics[width=0.48\textwidth]{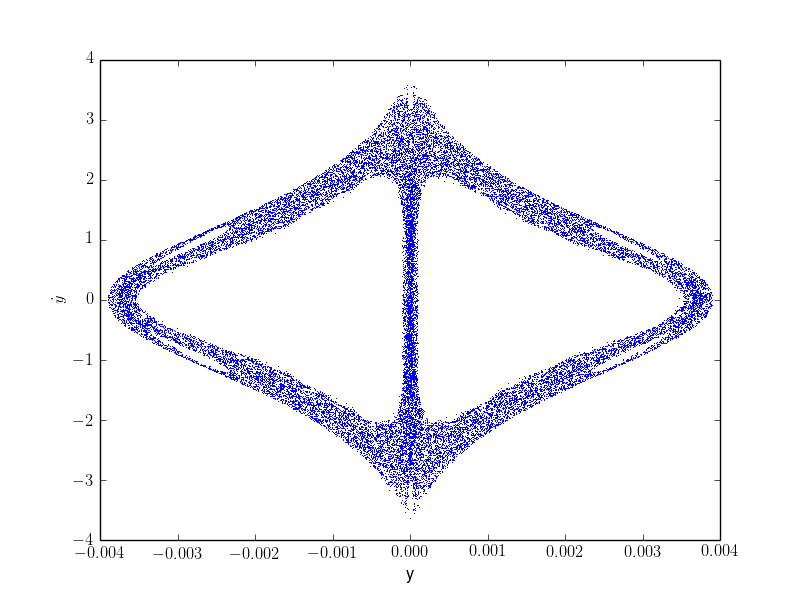}
  \caption{Orbit (left) and two-sided Poincar\'e SOS plot (right) for
    an initially mildly eccentric orbit with $(I_1, I_2)=(0.0002,
    0.004)$ in the chaotic zone predicted by the nKAM method.  This
    orbit shows a broad stochastic sheath with significant turning
    point variation although the orbit \emph{does} remain
    isolated.
    \label{fig:orb94} }
\end{figure*}

The nKAM method does appear to underpredict the stochastic regions.
Indeed, many of the orbits with $I_1<0.001$ and $0.010<I_2<0.015$ that
show larger values of Lyapunov exponent in Fig. \ref{fig:lyapdens}
exhibit the qualitative signature of chaos, although the nKAM method
is able to construct a torus.  Choosing a specific example from the
$\epsilon=0.3$ model, Fig. \ref{fig:orb130} describes an initially
nearly circular orbit with $(I_1, I_2)=(0.0004, 0.011)$, which is near
the middle of the higher Lyapunov value band near the corotation
resonance.  The origin of this discrepancy is not clear.  Increasing
the number of nodes in Fourier series by factors of 4 and 16 do not
change the loci of broken tori significantly.  More likely, as in the
standard application of the KAM theorem, the quadratically convergent
algorithm appears to robustly find tori in the vicinity of broken tori
whenever possible.  That is, away from the vanishing denominators, the
iterative algorithm tends toward a solution if at all possible.  This
suggests that the nKAM procedure is \emph{conservative}: it will find
regular orbits if they are in a small neighbourhood of the specified
initial condition even if irregular orbits can be found there as well.
Conversely, the Lyapunov exponent procedure seems to find both regular
and irregular orbits in the same neighbourhood, if the both exist, and
possibly switches between them (more on this below).

A second limitation of the nKAM procedure using the fundamentally
linearised solution of the HJ equation is its intrinsically
perturbative nature.  So, it is possible that nKAM method has been
pushed beyond its domain of accuracy for this problem resulting in the
underprediction of stochasticity in this case.  For larger value of
$I_1$, the nKAM appears to be a sensitive indicator of weak or
confined \citep[e.g.][]{Dvorak.etal:1998,Winter.etal:2010} chaos,
perhaps as expected, than the Lyapunov exponent value.  However, the
different characterisations persist even for $\epsilon\lesssim0.2$,
and therefore, so strong non-linearity is not the full explanation.

For another example, Fig. \ref{fig:orb94} shows the same plots as
Fig. \ref{fig:test_orbits} for a chaotic orbit in the ILR region.
Upon increasing $I_2$ at fixed $I_1$ beyond the locus of
nKAM-predicted broken tori, the SOS continues to exhibit significant
width, suggesting that the nKAM method underestimates the measure of
broken tori here as well.  Decreasing the amplitude of the
perturbation reveals the bifurcations and islands expected of a
stochastic layer; the trajectory appears regular over short times but
can switch between different regular morphologies.  This reinforces
our previous interpretation for the discrepancy between the apparently
small values of Lyapunov exponent when the orbit appears irregular
overall: the stickiness in the stochastic sheath results in regular
behaviour punctuated by jumps to different parent resonances near
islands.  One may easily verify there is clear structure in the SOS
over short finite-time periods that blends together over long periods.

\section{Summary and discussion}

\label{sec:summary}

\subsection{The method}

\label{sec:sum_method}

In this paper, we using the numerical KAM (nKAM) method described in
\citet{Weinberg:15a} to explore the location and frequency of tori
broken by a Galactic-bar perturbation.  In essence, this technique
predicts the existence of regular orbits (which have three actions)
and irregular orbits (with fewer than three actions) by an iterative
solution of the linearised Hamilton-Jacobi (HJ) equation.  The HJ
equation defines a canonical transformation to the action-angle
variable system, and we assume that a consistent solution for the
canonical formation implies existence and vice versa.  In section
\ref{sec:results}, we show that the nKAM computation yields useful
predictions predictions for regions of irregularity, however, the
method underpredicts the width of the chaotic zones for strong bars.
This trend may be understood as follows.  The nKAM method appears to
find a torus if one exists in the neighbourhood of the initial
conditions, even if broken tori exist in this same neighbourhood.
Conversely, the Lyapunov method tends to find exponential divergence
in the same circumstances, if it exists, although orbits may switch
behaviour suddenly, e.g. if the initially diffusing trajectory
encounters a regular island.  These differences are not expected, but
further elucidate that chaos is a broad catch-all term for orbits with
a wide variety of irregular morphology.

Thus, the two approaches reveal different information about the same
problem. The nKAM provides dynamical details about the principle and
secondary resonances for broken tori although the method is less
accurate for strong perturbations owing to its perturbative nature.
The Lyapunov analysis works for any perturbation, but the dynamical
origin of a positive Lyapunov exponent is not readily available.
Moreover, our investigation suggests that Lyapunov exponent analysis
does not provide useful information about stochastic orbits that
appear isolated to stochastic layers.  The morphology of the orbits
within these layers often exhibit large variance in orbital
frequencies while yielding the anomalously small Lyapunov exponents
typical of weak chaos.  Since the morphology of orbits as measured by
their time-averaged density is key to understanding the evolution of
patterns such as arms and bars, an random switching between families
matters.  This suggests that Lyapunov analysis may have less value for
some important galactic dynamics problems.  In addition, the nKAM is
capable of providing a new action-angle coordinate system for those
tori that survive a perturbation, although we do not exploit this
feature in the present paper.  Conversely, the nKAM method does not
provide a direct assessment of the magnitude of the chaos without
additional analysis; e.g. strong chaos may be inferred from the
relative magnitude of coefficients in $\varpi_{\bl}$ using the
\citet{Chirikov:79} criterion.

Our conclusion from this work is that galactic evolution under the
influence of strong patterns (bars, spiral arms, etc.) has an
important chaotic component.  As described in section
\ref{sec:intro}, this conclusion has been reached by previous
researchers as well.  Although standard perturbation theory provides
some physical insight, it is necessarily an incomplete picture of the
underlying dynamics.  We have some hope that the extended perturbation
theory described above may provide a bridge between analytic methods
and unencumbered n-body simulation.  Numerically, much of our
motivation for criteria for orbit integration stems from the study of
regular orbits.

\subsection{Implication for bars}

\label{sec:sum_bars}

In addition to characterising the new method, our main goal is a study
of the degree of stochasticity induced by a stellar bar in an
initially axisymmetric galaxy disc.  The importance of weak chaos in
structuring stable orbits under non-axisymmetric perturbations and the
underlying dynamics is nicely illustrated by
\citet{Romero-Gomez.etal:2006, Romero-Gomez.etal:2007,
  Athanassoula2009} who describe the importance of weak chaos in
creating rings in barred galaxies.  Their dynamical description of
orbit transitions at unstable points is essentially that described in
section \ref{sec:results} (Fig. \ref{fig:test_orbits}d and associated
discussion.).  There are, of course, other approaches.  For example,
\citet{Bountis.etal:2012} describe a probabilistic approach that
simultaneously discriminates between weak and strong chaos.  An
advantage of the nKAM approach is that it provides a numerical tool
for a wholesale investigation of phase space as a in a computationally
efficient way.  At the same time, it produces a multiple-resonance
model Hamiltonian to further investigate broken tori.

Based on an action-space expansion, our nKAM investigation provides a
provides a statistical view of irregularity caused by a perturbation
or pattern.  We assumed an analytic bar model and examined the
influence of its quadrupole moment on the disc.  The bar here is
modelled by a homogeneous ellipsoid with the axis ratio and density
profile typical of observed bars (see section \ref{sec:bar}).  The bar
does not evolve with time or is it self-consistent in any way; this is
an idealisation that makes this present investigation tractable,
although extensions with more generality are possible.  The key
findings are as follows:
\begin{enumerate}
\item The principle resonances causing broken tori in the inner galaxy
  are the corotation $(0, 2)$ and the OLR $(1, 2)$ resonances.  The
  secondary resonances with the largest amplitude for corotation are
  ILR $(-1, 2)$ and OLR $(1, 2)$; their amplitude relative to the
  primary resonance is roughly 3\% for a strong bar.  The
  strongest-amplitude secondary resonance for OLR is $(2, 2)$; their
  amplitude relative to the primary is roughly 10\%.
\item It may seem strange that major low-order resonances affect each
  other.  However, for a strong bar perturbation, i.e. when the
  non-axisymmetric force is similar in magnitude to the axisymmetric
  force, the width of libration zones for these resonances are
  comparable to the actions themselves.
\item As the amplitude increases, approaching that of a strong bar,
  many orbits in the vicinity of bar become chaotic, leaving a small
  phase-space region of modestly eccentric non-chaotic orbits between
  the ILR and corotation resonances.  This the same part of phase
  space responsible for supporting the bar itself
  \citep[e.g.][]{Contopoulos1980}.  This suggests that bar strength
  may be limited by the chaos induced by the bar itself, preventing
  further growth.
\item The Lyapunov exponent values do not accurately diagnose the
  chaos caused by the stochasticity sheaths that form around primary
  resonances.  The use of these exponents may provide a biased
  assessment of irregularity for galactic dynamics.
\end{enumerate}

\subsection{Implications for n-body simulations}

As described by \citet{Chirikov:79}, in \citetalias[][section
4.1]{Weinberg:15a}, and in the previous sections of this paper,
non-deterministic behaviour owing to mutual interaction of resonances
gives rise to a \emph{sheath} of chaotic behaviour around the primary
resonance.  In perturbation theory based on the averaging principle,
we isolate the degree of freedom corresponding to a resonance by
examining its dynamics close to its (unstable) homoclinic trajectory.
The phase frequency corresponding to this degree of freedom will be
small and we call the corresponding action the \emph{slow} action.
The slow action corresponding to this homoclinic trajectory of the
primary resonance will be, in general, a linear combination of the
actions defined by the separable coordinates of the unperturbed
background potential.  Therefore, motion in this sheath can result in
changes of the original actions of order the width of the libration
zone.

For the specific example explored in section \ref{sec:results},
chaotically induced switching of orbit families occurs in regions of
very small phase-space volume near unstable points.  The pervasiveness
of this fine-scale structure in phase space induced by chaos motivates
a technical campaign to determine sufficient conditions on n-body
simulation necessary to accurately recover the important consequences
of the mixed regular and irregular dynamics of galaxies.  Typical
time-step selection algorithms for n-body simulations choose a
fraction of the rate of change of position or velocity of a trajectory
in its gravitational field.  The scale of the gravitational field may
be determined by examining the derivative of forces or from prior
knowledge about the desired resolution length scale.  However, the
dynamics of interacting resonances is sub-dimensional.  Therefore, the
details of the dynamics around the primary resonance may require the
resolution of scales much smaller than those characteristic of the
trajectory as a whole. For example, if a single step in the ODE solver
moves the orbit through the chaotic zone altogether, the chaotic
dynamics may disappear from the simulation.  Indeed, accurate
computation of the Lyapunov exponents in \citetalias[][section
3.2]{Weinberg:15a} suggest that the required time steps are much
smaller than those chosen for n-body simulations; larger time steps
resulted in negative rather than positive exponents in many cases.
Similar issues were seen in the computation of SOS plots for section
\ref{sec:results}.  The topology of the SOS plots for broken tori are
often not converged until time steps are several orders of magnitude
smaller than the characteristic orbital times for the RK4 time
stepper.

\subsection{Future work}
\label{sec:future}

Many of our analytic tools for studying modes and secular processes
excited by bars and strong spiral patterns are based on the existence
of regular orbits.  Our success in studying the statistical
implication of an imposed pattern as described in section
\ref{sec:sum_bars} leads to a secondary longer-term goal: can we
extend perturbation theory methods to understand and predict the
long-term future of secularly evolving galaxies that includes the
effects of chaos?

Indeed, an immediate consequence of the unbroken tori found using nKAM
is a new basis for secular perturbation theory.  The framework
described in \citetalias{Weinberg:15a} and in section \ref{sec:KAM}
provides the new regular trajectories in an action-angle
representation when they exist.  This basis may be used to estimate
bar kinematic signatures, perform stability analyses, and so on; any
task previously performed using Hamiltonian perturbation theory may be
performed using the new action-angle representation.  This naturally
allows the standard tools of secular perturbation theory
\citep[i.e.][]{Tremaine.Weinberg:84} to be used in the mixed context
of regular and irregular orbits.  Of course, this only makes sense if
many of the tori remain unbroken.  Based on the results of section
\ref{sec:results}, this roughly true for all but the strongest bars.

The orbit morphology in the perturbed regions may be significantly
different than the unperturbed one.  For example,
Fig. \ref{fig:regular_orb} describes an regular orbit reconstructed by
the nKAM method.  The SOS plot [Fig. \ref{fig:chaotic_sos}] reveals
that the simple rosette orbit has been destroyed and replaced with an
orbit whose turning points are trapped by the rotating bar potential.
These changed morphologies of nKAM-determined regular orbits are
capable of supporting new classes of responses not possible with the
original unperturbed regular system.  For Fig. \ref{fig:chaotic_orb},
the new orbit may still be bar supporting, but less so than a
classical x1 bar orbit from \citet{Contopoulos1980}.  In addition,
each new class of regular orbits is likely to present a different
kinematic signature than the original unperturbed orbit, and an
ensemble chosen to represent the bar may be used to predict kinematic
signatures.

A companion paper applies the nKAM method to a three-dimensional but
axisymmetric system.  A natural extension of this work to three
dimensions would explore the role stochasticity in coupling between
the planar and vertical degrees of freedom.  For some examples of the
questions to be addressed, does the bar perturbation drive disc
thickening through stochastic means?  Do broken tori play a role in
producing a pseudo-bulge?  As described in \citetalias{Weinberg:15c},
the fully asymmetric three-dimensional implementation of the nKAM
method is computational challenging but surmountable.  Each torus is
represented by several three-dimensional Fourier series describing the
canonical generating function and tables of the unperturbed orbits as
a function of angles. The numerical partial derivatives required by
the nKAM method requires that three-dimensional action cube be finely
sampled by these tori.  This leads to a large RAM requirement.  The
current code parallelises torus computation and shares the results
with all processes.  A redesigned code would use an action-space
domain decomposition to eliminate need for each compute node have a
full copy of the Fourier series and angles grids for each torus.  This
will allow a computationally feasible, general, three-dimensional nKAM
code.

The example here assumes the simplest quasi-periodic time-dependent
perturbation: one with a single constant pattern speed.  As described
in \citetalias{Weinberg:15a} and sections above, the nKAM method may
be extended to include general time-dependent perturbations by
breaking down the time-dependent perturbation using a Fourier (or
Laplace) series approximation.  A related question is the effect of
the slow evolution of the bar and background gravitational field owing
to secular torques.  For example, does the slow evolution increase or
decrease the size and effects of the stochastic layers?  Naively, the
effective frequency of the slow evolution is much smaller than the
orbital frequencies in the inner galaxy, and this makes the coupling
between the natural frequencies and the evolution frequencies weak.
Conversely, the low spatial frequency of the secular evolution may
effectively couple to the primary perturbation for a strong bar,
providing new channels for separatrix breaking.  Moreover, the
existence of chaos in the vicinity of orbits supporting the bar is
likely to affect capture into libration.  For example, since bars are
formed by capture into libration, we might expect diffusion in and out
of the libration zone as the bar pattern speed changes.  Results from
section \ref{sec:results} suggest that stochastic regions caused by
the forming and growing bar increase with bar strength.  Understanding
influence of slow evolution on stochasticity from first principles and
the possibility that bars are chaotically self-limited is the next
challenge.

\section*{Acknowledgments}

This work was supported in part by NSF awards AST-0907951 and
AST-1009652.  MDW gratefully acknowledges support from the Institute
for Advanced Study, where work on this project began.  Many thanks
Douglas Heggie for many valuable comments on an early version of this
manuscript.

\bibliographystyle{mn2e}

\label{lastpage}
\null

\end{document}